\documentclass{article}
\usepackage{setspace}
\usepackage{braket}
\usepackage{bbm}
\usepackage{tcolorbox}
\usepackage{graphicx}
\usepackage{amsthm}
\usepackage{multicol}
\usepackage{hyperref}
\hypersetup{
    colorlinks,
    citecolor= orange,
    filecolor= blue,
    linkcolor= blue,
    urlcolor= green
}
\usepackage{mathrsfs}
\usepackage{appendix}
\usepackage{amssymb}
\usepackage{relsize}
\usepackage{amsthm}
\usepackage{ amssymb }
\usepackage{amsmath}
\usepackage[utf8]{inputenc}
\usepackage[english]{babel}
\usepackage[a4paper, total={6in, 9.2in}]{geometry}
\usepackage{amsmath}

\newtheorem{definition}{Definition}
\newtheorem{definition2}{Lemma}

\newtheorem{definition4}{Theorem}
\newtheorem{definition5}{Remark}
\newtheorem{definition6}{Claim}

\newtheorem{Co}{Corollary}
\usepackage[T1]{fontenc}
\usepackage{lmodern}

\usepackage{tikz-cd}

\begin{titlepage}

\title{Finite-time quantum equilibration for continuous variables}
\author{Alberto Acevedo, Antonio Falcó \\ 
Departamento de Matemáticas, Física y Ciencias Tecnológicas \\
Universidad Cardenal Herrera-CEU, CEU Universities,\\ 
Calle San Bartolome, 55, Alfara del Patriarca 46115, Valencia, Spain}
\date{}

\end{titlepage}

\begin{document}

\thispagestyle{plain}

\maketitle
\begin{abstract}
Leveraging the techniques found in the literature on Quantum Equilibration for finite dimensional systems, we develop the theory of Quantum Equilibration
for the case of infinite dimensional systems, particularly the cases where the dynamics-generating Hamiltonians have continuous spectrum. The main goal of this paper will be to propose a framework to extend the results obtained by Short in
\cite{short}, 
where estimates for the equilibration-on-average and effective equilibration for the case of Hamiltonians with continuous spectrum are derived. We will show that in the latter setting, it is compulsory to constrain ourselves to finite time equilibration; we then develop estimates analogous to the main results in the proposed setting. 
\end{abstract}

\begin{section}{Introduction}
\;\;\; Due to the recurrence and time reversal invariance properties of finite dimensional quantum systems undergoing unitary evolution, equilibration in the
sense of Boltzmann’s H-Theorem, which implies that entropy can only grow over time, is impossible \cite{haar2}. In recent times, this apparent contradiction between quantum theory and thermodynamics has been one of the main focal points of research; for a historical and theoretical overview the reader is invited to see \cite{gogolin}. The proposed solution to this conundrum, which we will refer to as the \emph{equilibration problem}, is an application of the notions of \emph{equilibration-on-average} and \emph{equilibration-during-time-intervals}. In \cite{gogolin} these are defined as follows. 

\begin{itemize}
\item \textbf{Equilibration-on-average:} A time dependent property is said to equilibrate on
average if its value is for most times during the evolution close to some equilibrium
value.
\item \textbf{Equilibration-during-time-intervals:}  A time dependent property is said to equilibrate during a (time) interval if its value is close to some equilibrium value for all times in that interval.
\end{itemize}
Such notions of equilibration have been applied successfully to resolve the equilibration problem; some of the most noteworthy papers in the literature are \cite{short} \cite{short2} \cite{tasaki}  \cite{Linden} \cite{ri1} \cite{ri2} \cite{ri3} \cite{hh} \cite{cra}. In this paper we will be primarily focused on developing a framework that parallels the work of Short et al \cite{short} \cite{short2} for the case of continuous variables.\\

In the work of Reimann \cite{ri1} \cite{ri2} it is shown that the expectation value of a quantum observable will equilibrate to
an approximately static value, given some weak assumptions about the Hamiltonian generating the dynamics and the initial state; namely that the Hamiltonian has non-degenerate spectral gaps. The latter is indeed a more technical description of equilibration-on-average which we will formalize in the following sections. Heuristically speaking, a quantum observable equilibrates with respect to some Hamiltonian $\mathbf{\hat{H}}=\sum_{i}\lambda_{i}\mathbf{\hat{P}}$ and initial quantum states $\boldsymbol{\hat{\rho}}_{0}$ if the statistics of said observable virtually never deviates from the statistics of the same observable with respect to the dephased state $\sum_{i}\mathbf{\hat{P}}_{i}\boldsymbol{\hat{\rho}}_{0}\mathbf{\hat{P}}_{i}$, i.e. the equilibrium state \cite{short}. In \cite{short} \cite{short2}, Reimann's primary result from \cite{ri1} is extended and generalized to include analogous results for subspaces and for finite equilibration time. To the extent that we are informed,  these papers, and all previous work on equilibration of this sort for that matter, focus only on the cases where the quantum mechanical systems in question are described by finite dimensional Hilbert spaces. There does not therefore seem to exists a framework that addresses the quantum equilibration problem for the case of quantum mechanical systems described by infinite dimensional Hilbert spaces and whose dynamics are characterized by Hamiltonian's have continuous spectrum.\\

For the finite dimensional case, the finiteness of the spectrum of the Hamiltonian generating the dynamics makes it relatively simple to obtain estimates for the equilibration of some specified observer given some initial states. However, when one deals with infinite dimensional Hilbert spaces one must now consider other types of spectrum such as absolutely continuous and singular continuous spectrum, together known as continuous spectrum, which make the notion of effective dimension \cite{short} \cite{short2} difficult to recapture. To understand what the notions of equilibration posed in \cite{ri1} \cite{ri2} \cite{short} \cite{short2} extend to, for the case of continuous variables, we must first consider a rigorous development of the relevant semigroup theory and the respective ergodic averages; this will be developed in section \ref{eqn:sec3} of this paper. We will see that the existence of a nontrivial ergodic average (for infinite time) depends on the spectral properties of the Hamiltonian. This will in turn force us to adopt a framework that introduces a finite equilibration time. In part section \ref{eqn:sec4} we will formally define equilibration averages and present the main results that we wish to extend, namely those of Short from the work \cite{short}. Finally, in section \ref{eqn:sec5} we address the problem of equilibration for the case where the Hamiltonian has purely continuous spectrum. We provide bounds analogous to those posed by Short in \cite{short} for the estimation of equilibration-on-average and effective equilibration. These bounds will not only depend on the spectral properties of the dynamics-generating Hamiltonian and the initial state, but will also depend on other parameters such as the resolution limit and the finite equilibration time; parameters which were not present in the finite dimensional case addressed in \cite{short} \cite{short2} \cite{Linden} \cite{ri1} \cite{ri2} \cite{ri3}. In the final part of this paper we present an example.
\end{section}
\begin{section}{Preliminaries}

\;\;\; Before getting to the heart of the matter we shall present some of the mathematical conventions and preliminary material that we will use. Readers familiar with the usual notational conventions of quantum theory, density operators, functional analysis and operator theory may skip this section with little hindrance. \\
\;\;\;We will generally focus on operators acting on Banach spaces and particularly on Hilbert spaces which we will denote $\mathscr{B}$ and $\mathscr{H}$ respectively. 
    The corresponding norm will be denoted by $\|\cdot\|$. The dual space of $\mathscr{H}$ (respectively,  $\mathscr{B})$ will be denoted by $\mathscr{H}^{*}$ (respectively,  $\mathscr{B}^{*}).$
    As it is common in quantum theory, we will denote the 
    the elements in the Hilbert space by kets, i.e. $\ket{\boldsymbol{\psi}}$, and the dual elements by bras, i.e. $\bra{\boldsymbol{\psi}}$. The inner product of two elements of a Hilbert space $\mathscr{H}$ will be denoted by the usual bra-ket convention, that is, $\langle \cdot|\cdot\rangle$. 
    
\subsection{Bounded operators and its limits with respect to the strong operator topology}

The set of bounded operators, that is, linear and continuous  maps from a Banach space $(\mathscr{B}_1,\|\cdot\|_{\mathscr{B}_1})$ to a Banach space $(\mathscr{B}_2,\|\cdot\|_{\mathscr{B}_2})$ will be 
denote by 
$$
L(\mathscr{B}_1,\mathscr{B}_2):=\Big\{\mathbf{\hat{A}}:\mathscr{B}_1\rightarrow \mathscr{B}_2: \mathbf{\hat{A}} \text{ linear and continuous }\Big\}.
$$
It is a linear vector space with the operator norm of the operators $\mathbf{\hat{A}} \in L(\mathscr{B}_1,\mathscr{B}_2)$, i.e. $ \|\mathbf{\hat{A}}\|_{\mathscr{B}_1 \to \mathscr{B}_2}$, is defined as follows.

\begin{definition}[\textbf{Operator Norm}]
\small
Consider an operator $\mathbf{\hat{A}} \in L(\mathscr{B}_1,\mathscr{B}_2).$  The operator norm is defined as follows. 
\begin{equation}
\|\mathbf{\hat{A}}\|_{\mathscr{B}_1 \to \mathscr{B}_2}:= \sup\Bigg\{\frac{\big\|\mathbf{\hat{A}}\big|\psi\big\rangle\big\|_{\mathscr{B}_{2}}}{\big\|\big|\psi\big\rangle\big\|_{\mathscr{B}_{2}}} : \big|\psi\big\rangle \neq 0 \;\; and \;\; \big|\psi\big\rangle \in \mathscr{B}_{1} \Bigg\}.
\end{equation}
\normalsize
\end{definition}
Observe that the operator norm is the smallest constant $C >0 $ such that $$\big\|\mathbf{\hat{A}}\big|\psi\big\rangle\big\|_{\mathscr{B}_{2}}\leq C\,\big\|\big|\psi\big\rangle\big\|_{\mathscr{B}_{1}}$$ for all $|\psi\rangle \in \mathscr{B}_{1}$. The operator norm is a norm on the vector space $L(\mathscr{B}_1,\mathscr{B}_2)$, and it is a Banach space with respect to this norm.
In this paper, we will encounter limits of the form $\lim_{t\rightarrow \infty} \mathbf{\hat{A}}_{t}$. When we write this, unless otherwise stated, the limit will be assumed to be taken with respect to the strong operator topology. We formally define this notion here below. 
\begin{definition}[\textbf{Operator limit with respect to the strong operator topology}]
\label{eqn:sot}
\small
Consider a set of operators $\big\{\mathbf{\hat{A}}_{t}\big\}_{t \in \mathbb{R}_+} \subset L(\mathscr{B}_1,\mathscr{B}_2).$ 
We say that $\mathbf{\hat{A}}_{t}$ converges to $\mathbf{\hat{A}}$ as $t\rightarrow \infty$ with respect to the strong operator topology if for every $\big|\psi\big\rangle \in \mathscr{B}_{1}$ it holds
\begin{equation}
\lim_{t\rightarrow \infty}\big\|\mathbf{\hat{A}}_{t}\big|\psi\big\rangle - \mathbf{\hat{A}}\big|\psi\big\rangle\big\|_{\mathscr{B}_{2}} = 0.
\end{equation}
Symbolically we will just write 
\begin{equation}
    \lim_{t\rightarrow \infty} \mathbf{\hat{A}}_{t} = \mathbf{\hat{A}}.
\end{equation}
\normalsize
\end{definition}

\subsection{Bochner integration}
\;\;\;Not only shall we be interested in the convergence of sequences of operators, we shall also be interested in the integration of operator-valued functions, in particular integrals defining the T-time average over the parameter $t$, namely $\langle\!\langle\mathbf{\hat{A}}_{t}\rangle\!\rangle_{T}:=\frac{1}{T}\int_{0}^{T}\mathbf{\hat{A}}_{t}dt$. To establish a mathematically rigorous interpretation of such an object we will use the notion of the Bochner integral \cite{boch}. The Bochner integral is usually introduced via a general abstract measure space construction, however, since we will only worry about integrals over the positive reals, we will avoid such abstractions and just define the Bochner integral accordingly. Given a function $f(t):[0,T)\rightarrow \mathscr{B}$, where $\mathscr{B}$ is a Banach space, one may approximate such a function, in the sense of the measure topology, with simple functions $s_{N}(t):=\sum_{n=1}^{N}\chi_{E_{n}}(t)\mathbf{\hat{b}}_{n}$; $E_{n}$ being disjoint open subsets of $[0,T)$ and $\mathbf{\hat{b}}_{n}$ elements of $\mathscr{B}$. Integration of the simple functions $s_{N}(t)$ is quite transparent, i.e. $\int_{0}^{T}s_{N}(t)dt = \sum_{n=1}^{N}\mu(E_{n})\mathbf{\hat{b}}_{n}$, whence determining whether or not one might arbitrarily approximate $\mathbf{\hat{A}}_{t}$ with simple functions is of great interest. The above discussion motivates the following definition. 
\begin{definition}[
\textbf{Bochner integrable}]
\label{eqn:boch}
Consider a set of operators $\{\mathbf{\hat{A}}_{t}\}_{t \in \mathbb{R}_+}$, where for all $t$ we have $\mathbf{\hat{A}}_{t}\subset L(\mathscr{B}_1,\mathscr{B}_2).$ If there exists a sequence of simple functions $s_{N}(t):= \sum_{n=1}^{N}\chi_{E_{n}}(t)\mathbf{\hat{b}}_{n}$, $\mathbf{\hat{b}}_{n}\in L(\mathscr{B}_1,\mathscr{B}_2)$  such that
\small
\begin{equation}
\lim_{N\rightarrow \infty}\int_{0}^{T}\|\mathbf{\hat{A}}_{t}-s_{N}(t)\|_{\mathscr{B}_{1}\rightarrow\mathscr{B}_{2}}dt  =0 
\end{equation}
\normalsize
i.e. convergence to zero wrt the measure topology, then we say that $\mathbf{\hat{A}}_{t}$ is Bochner integrable. The respective Bochner integral is then defined as follows
\small
\begin{equation}
\int_{0}^{T}\mathbf{\hat{A}}_{t}dt = \lim_{N\rightarrow \infty}\int_{0}^{T}s_{N}(t)dt
\end{equation}
\normalsize
\end{definition}
It turns out that it is quite easy to prove that a function is Bochner integrable; to that end, we have the following result
\begin{definition5}[\textbf{Proving Bochner integrability}]
Let $\mathbf{\hat{A}}_{t}$ a be a function from $[0,T)$ to Banach space $\mathscr{B}$. Then, $\mathbf{\hat{A}}$ is Bochner integrable over $[0,T)$ if and only if 
\begin{equation}
\int_{0}^{T}\big\|\mathbf{\hat{A}}_{t}\big\|dt <\infty   
\end{equation}
\end{definition5}
With the above tools, we will now be able to make sense of limits of the form $\lim_{t\rightarrow\infty}\langle\!\langle\mathbf{\hat{A}}_{t}\rangle\!\rangle_{T}$; these will play an important role in the following sections. 
\subsection{$c_{0}$-semigroups}
\;\;\; In this work we will primarily be interested in the so-called \emph{ergodic averages}, formally defined in the following section, on strongly continuous one-parameter semigroups, aka $c_{0}$-semigroups (we will often just call them continuous semigroups for short). Here, we give a definition of continuous semigroups and provide the milestone-result by Stone which characterizes unitary semigroups. 
\begin{definition}[$c_{0}$-\textbf{semigroup}]
\label{eqn:semi}
\small
A strongly continuous one-parameter semigroup on a Banach space $\mathscr{B}$ is a map $\mathbf{\hat{T}}_{t}:\mathbb{R}_{+}\rightarrow L\big(\mathscr{B}\big)$(where $L\big(\mathscr{B}\big)$ is the space of bounded operators on $\mathscr{B}$) such that 
\begin{itemize}
    \item $\mathbf{\hat{T}}_{0} = \mathbf{\hat{I}}$, (the identity operator on $\mathscr{B}$).
    \item $\forall t,s \geq 0: \mathbf{\hat{T}}_{t+s} = \mathbf{\hat{T}}_{t}\mathbf{\hat{T}}_{s}$
    \item $\forall \mathbf{\hat{v}} \in \mathscr{B} : \big\|\mathbf{\hat{T}}_{t}\mathbf{\hat{v}}- \mathbf{\hat{v}}\big\|\rightarrow 0$, as $t\rightarrow 0^{+}$
\end{itemize}
\normalsize
\end{definition}
For unitary $c_{0}$-semigroups Stones's theorem \cite{Simon} stablishes a one to one correspondence between self-adjoint operators and strongly continuous one-parameter unitary semigroups; we present this for the case of a semigroup acting on a Hilbert space below. 
\small
\begin{definition4}[\textbf{Stone's Theorem}]
\label{eqn:stone}
Consider a strongly continuous unitary one-parameter semigroup on a Hilbert space $\mathscr{H}$. Namely, $\mathbf{\hat{U}}_{t} :\mathbb{R}_{+}\rightarrow L\big(\mathscr{H}\big)$.
Then there exists a unique (possibly unbounded) operator $\mathbf{\hat{A}} : \mathrm{Dom}\big(\mathbf{\hat{A}}\big)\rightarrow \mathscr{H}$ that is self-adjoint on $\mathrm{Dom}\big(\mathbf{\hat{A}}\big)$ such that 
$\forall t \in \mathbb{R}_{+}: \;\; \mathbf{\hat{U}}_{t} = e^{it\mathbf{\hat{A}}}$
where
\begin{equation}
    \mathrm{Dom}\big(\mathbf{\hat{A}}\big) = \Bigg\{\big|\psi\big\rangle \in \mathscr{H}\Bigg| \lim_{\varepsilon \rightarrow 0} \frac{-i}{\varepsilon}\big(\mathbf{\hat{U}}_{\varepsilon}\big|\psi\big\rangle-\big|\psi\big\rangle\big)\;\; exists \;\;\Bigg\}
\end{equation} 
Conversely, let $\mathbf{\hat{A}}:\mathrm{Dom}\big(\mathbf{\hat{A}}\big)\rightarrow \mathscr{H}$ be a (possibly unbounded) self-adjoint operator on $\mathrm{Dom}\big(\mathbf{\hat{A}}\big) \subset \mathscr{H}$. Then, the one-parameter family $\big\{\mathbf{\hat{U}}_{t}\big\}_{t\in\mathbb{R}_{+}}$ of unitary operators defined by $\forall t\in \mathbb{R}_{+}: \;\; \mathbf{\hat{U}}_{t} := e^{it\mathbf{\hat{A}}}$ is a strongly continuous one-parameter semigroup. 
\end{definition4}
\normalsize
\subsection{The spectrum of an operator}
\;\;\; Understanding the dynamics generated by a unitary semigroup will require an understanding of the spectral properties of the generator of said semigroup. For the case where $di(\mathscr{H})=\infty$, the generator $\mathbf{\hat{A}}$ may have different types of spectrum. In standard books on functional analysis and/or operator theory \cite{Simon} \cite{lax} \cite{retherford}, one usually sees the following definition of the spectrum of an operator.
\begin{definition}[\textbf{The Spectrum of an operator}]
\label{eqn:thespectrum}
Let $\mathbf{\hat{H}}$ be an arbitrary self-adjoint operator acting over a Hilbert space $\mathscr{H}$; the spectrum of said operator, denoted $\sigma(\mathbf{\hat{H}})$, is the union of following disjoint sets.
\begin{itemize}
\item \textbf{point spectrum} of $\mathbf{\hat{H}}:= \overline{\sigma}_{pp}( \mathbf{\hat{H}})$ : The closure of the set of eigenvalues of $\mathbf{\hat{H}}$, i.e. $\sigma_{pp}(\mathbf{\hat{H}} )$ is the set of eigenvalues of $\mathbf{\hat{H}}$ (this is called the pure point spectrum \cite{Simon}). 
\item \textbf{continuous spectrum} of  $\mathbf{\hat{H}}:= \sigma_{c}( \mathbf{\hat{H}})$ : 
 Consists of all scalars, $\lambda$ that are not eigenvalues but make the range of $\mathbf{\hat{H}}-\lambda\mathbf{\hat{I}}$ a proper dense subset of $\mathscr{H}$.  
 \item \textbf{residual spectrum} of $\mathbf{\hat{H}}:= \sigma_{r}(\boldsymbol{\hat{H}})$: $\mathbf{\hat{H}}-\mathbf{\hat{I}}$ is injective but does not have dense range.
\end{itemize}
\begin{equation}
\sigma(\mathbf{\hat{H}}) = \bar{\sigma_{pp}}(\mathbf{\hat{H}} )\cup \sigma_{c}( \mathbf{\hat{H}})\cup  \sigma_{r}( \mathbf{\hat{H}})
\end{equation}
\end{definition}
Note that for the case were $\mathbf{\hat{H}}$ is self-adjoint $\sigma_{r}(\mathbf{\hat{H}})$ is empty. Hence, for self-adjoint $\mathbf{\hat{H}}$, $\sigma\big(\mathbf{\hat{H}}\big) = \overline{\sigma}_{pp}(\mathbf{\hat{H}})\cup\sigma_{c}(\mathbf{\hat{H}})$.

\subsection{Trace-class and Hilbert-Schmidt operators}

\;\;\; Let us now introduce the notion of trace-class and Hilbert-Schmidt operators. These are families of operators that are of great importance in the study of quantum mechanics. Let $(\mathscr{H}, \|\cdot\|_{\mathscr{H}})$ be a separable complex Hilbert space. We will write $L( \mathscr{H} ) = L(\mathscr{H},\mathscr{H})$ and $\|\cdot\| = \|\cdot\|_{\mathscr{H} \to \mathscr{H}}$ for simplicity. Next, we define the \emph{trace of an operator}.
\begin{definition}[\textbf{Trace of an operator}
] Let $\{\big|\psi_{n}\rangle\}_{n}$ be any orthonormal basis of $\mathscr{H}$. Given $\boldsymbol{\hat{A}} \in L( \mathscr{H} )$,  the trace of $\boldsymbol{\hat{\sigma}}$ is defined as follows
\begin{equation}
\mathrm{Tr}\{\boldsymbol{\hat{A}}\} := \sum_{n}\big\langle \psi_{n}\big|\boldsymbol{\hat{\sigma}}\big|\psi_{n}\big\rangle.
\end{equation}
The value of the trace, assuming that it exists, is independent of the basis chosen. 
\end{definition}

The latter allows us to introduce the notion of a Hilbert-Schmidt operator and a trace-class operator. 

\begin{definition}[\textbf{Hilbert-Schmidt and Trace-Class operators}]
\label{eqn:hilsch}
Let $\boldsymbol{\hat{A}} \in L( \mathscr{H} )$, then the Hilbert-Schmidt norm and Trace norm of $\boldsymbol{\hat{A}}$ are respectively the following
\begin{equation}
\|\boldsymbol{\hat{A}}\|_{H.S.}:=\sqrt{\mathrm{Tr}\big\{\boldsymbol{\hat{A}}^{\dagger}\boldsymbol{\hat{A}}\big\}}.
\end{equation}
An operator 
\begin{equation}
\|\boldsymbol{\hat{A}}\|_{1}:=\mathrm{Tr}\Big\{\sqrt{\boldsymbol{\hat{A}}^{\dagger}\boldsymbol{\hat{A}}}\Big\}<\infty.
\end{equation}
We denote the set of Hilbert-Schmidt operators and the Trace-Class operators respectively by $\mathscr{S}_2( \mathscr{H})$, $\mathscr{S}_1( \mathscr{H})$.
\end{definition}
An interesting class of operators are the finite rank operators. These are operators that can be written as a finite sum of rank one operators as follows. Considering the tensor product of two vectors $\ket{\phi}$ in $\mathscr{H}$
and $\bra{\psi} \in \mathscr{H}^{*}$ i.e. $\ket{\phi} \otimes \bra{\psi}:=|\psi\rangle\langle\phi|$ we can construct the following algebraic tensor product 
space 
\begin{equation}
\mathscr{H}\otimes \mathscr{H}^{*}:=\Bigg\{\sum_{i}\ket{\phi_{i}}\otimes \bra{\psi_{i}}: \ket{\phi_{i}} \in \mathscr{H} \;\; and \;\; \bra{\psi_{i}} \in \mathscr{H}^{*} \Bigg\}.
\end{equation}
We can identify each element of $\mathscr{H}\otimes \mathscr{H}^{*}$ with a linear operator acting on $\mathscr{H}$, i.e. $\ket{\phi}\otimes \bra{\psi} \in \mathscr{H}\otimes \mathscr{H}^{*}$ is identified with the operator $\boldsymbol{\hat{A}}_{\ket{\phi}\otimes \bra{\psi}}$ defined by $\boldsymbol{\hat{A}}_{\ket{\phi}\otimes \bra{\psi}}\ket{\varphi}:\langle\psi|\varphi\rangle \ket{\phi}$. The set of finite rank operators acting on $\mathscr{H}$ is then defined by
$F(\mathscr{H}):= \mathscr{H}\otimes \mathscr{H}^{*}$, and it is a linear subspace of $L(\mathscr{H})$.
It is useful to note that 
    \begin{equation}
    F(\mathscr{H})\subseteq \mathscr{S}_{1}( \mathscr{H} )\subseteq\mathscr{S}_{2}( \mathscr{H}) \subseteq L( \mathscr{H}) 
    \end{equation}
Moreover, it is known that $\mathscr{S}_2(\mathscr{H}) = \overline{\mathscr{H}\otimes \mathscr{H}^{*}}^{\|\cdot\|_{HS}}$ is the closure of the finite rank operators with respect to the Hilbert-Schmidt norm. Furthermore, owing to the equality $$\|\ket{\phi}\bra{\psi}\|_{HS} = \|\ket{\phi}\|\|\ket{\psi}\|,$$
the Hilbert-Schmidt norm is a \emph{cross-norm}, whence $\mathscr{S}_2(\mathscr{H})$ is a tensor product Hilbert space \cite{Hackbusch}. In a similar fashion, the trace-class operators are a tensor product Banach space taking with respect to the injective norm
$\|\cdot\|_{\vee(\mathcal{H},\mathcal{H}^{*})}$ defined in the \emph{algebraic tensor space} $\mathscr{H}\otimes \mathscr{H}^{*}$ as follows
\begin{equation}
    \|\boldsymbol{\hat{A}}\|_{\vee(V,W)} = \sup\left\{|(\bra{\varphi} \otimes \ket{\psi})(\boldsymbol{\hat{A}})|: \bra{\varphi} \in \mathscr{H}^*,\, \| \bra{\varphi}\|_{{\mathcal{H}}^*}=1\, ; \ket{\psi} \in \mathscr{H}\, \|\ket{\psi}\|_{\mathscr{H}}=1 \right\},
\end{equation}
where $(\bra{\varphi} \otimes \ket{\psi})(\ket{\eta} \otimes \bra{\mu}):= \langle\varphi|\eta\rangle \overline{\langle\mu|\psi\rangle}.$ Thus, $\mathscr{S}_1(\mathscr{H}) = \overline{\mathscr{H}\otimes \mathscr{H}^{*}}^{\|\cdot\|_{\vee(\mathcal{H},\mathcal{H}^{*})}}$ is a tensor product Banach space \cite{Hackbusch}.

Indeed, the space of trace class operators, and the space of Hilbert-Schmidt operators are families of compact operators \cite{Simon} \cite{lax}. However, we will not need the concept of a compact operator in any greater generality than that of trace class and Hilbert-Schmidt operators; therefore, we omit a formal discussion of such a class of operators here. The definitions above make transparent the well-definedness of the trace when acting on density operators, i.e. as a map that will be restricted to the subspace $\mathscr{S}_{1}(\mathscr{H})$ of the larger Banach space $L(\mathscr{H})$. For the finite-dimensional case,  the trace trivially  exists; i.e. sum the elements of the diagonal. For the infinite-dimensional case one requires that the infinite sum of the singular values of the operator in question converges; this can be had, and it is indeed how the trace-class family is defined above. The density operators are positive operators whose eigenvalues/singular values sum to one by definition, whence the trace is well defined on this set. It turns out that density operators are also Hilbert-Schmidt operators \cite{han}, this is partially why we included a definition of Hilbert-Schmidt operators above. 
\subsection{The Density Operators}
We formally introduce the following class of operators, previously alluded to, that will be of great importance in the following sections.
\begin{definition}[\textbf{Density Operator}]
\label{eqn:densopdef}
Let $\mathcal{S}(\mathscr{H})$ be the set of positive-semidefinite operators with trace one acting in a Hilbert space $\mathscr{H}$. $\mathcal{S}(\mathscr{H})$ forms a convex subset of the respective Banach algebra $L(\mathscr{H})$ that they live in. Positive-semidefinite here means that given some operator $\boldsymbol{\hat{\rho}} \in \mathcal{S}(\mathscr{H})$, and for any $|\psi\rangle \in \mathscr{H}$, $\langle \psi|\boldsymbol{\hat{\rho}}|\psi\rangle\geq 0$; zero being obtained when the vector $|\psi\rangle$ is not in the support of $\boldsymbol{\hat{\rho}}$. We will refer to $\mathcal{S}(\mathscr{H})$ as the set of density operators and refer to any $\boldsymbol{\hat{\rho}}\in\mathcal{S}(\mathscr{H})$ as a density operator.
\end{definition}
Assuming that we have a quantum system in a state $\boldsymbol{\hat{\rho}} \in \mathsf{S}(\mathscr{H})$, we may compute its expectation value for an arbitrary observable $\mathbf{\hat{A}}$ as follows
    \begin{equation}
    \label{eqn: expectval}
    \big\langle \mathbf{\hat{A}}\big\rangle_{\boldsymbol{\hat{\rho}}} := \mathrm{Tr}\big\{\boldsymbol{\hat{\rho}} \mathbf{\hat{A}} \big\}.
    \end{equation}
Since the density operators are trace-class operators, and $\mathcal{S}(\mathscr{H})$ is a tensor Banach space, we can define for each density operator $\boldsymbol{\hat{\rho}}$ its tensor rank \cite{Hackbusch}. In particular, the set of density operators with tensor rank equal to one is given 
by 
 \begin{equation}
 \mathsf{S}_1\big(\mathscr{H} \big) := \big\{
    \boldsymbol{\hat{\rho}} \in \mathsf{S}\big(\mathscr{H} \big): \boldsymbol{\hat{\rho}} = \ket{\psi}\bra{\psi} \text{ for some } \ket{\psi} \in \mathscr{H} \text{ where } \|\ket{\psi} \| = 1
    \big\}
 \end{equation}
 and its elements are called \emph{pure states}. The set of density operators with tensor rank greater than one is called the set of \emph{mixed states}.  
If a density operator is not pure, i.e. mixed, it will be a convex mixture of density operators $\boldsymbol{\hat{\rho}} = \sum_{n}p_{n}\boldsymbol{\hat{\rho}}_{n}$, \:$\boldsymbol{\hat{\rho}}_{n} \in \mathsf{S}(\mathscr{H})$  and $\sum_{n}p_{n} = 1$. Mixed states of course live in the space $\mathsf{S}( \mathscr{H}) \setminus \mathsf{S}_1( \mathscr{H})$ owing to their projective properties. 
A pure state $\boldsymbol{\hat{\rho}}$ will satisfy the equality $\mathrm{Tr}\{ \boldsymbol{\hat{\rho}}^{2}\} =\mathrm{Tr}\{ \boldsymbol{\hat{\rho}}\} = 1;$ on the other hand, for a mixed state, $\mathrm{Tr}\{\boldsymbol{\hat{\rho}^{2}} \}< 1$ \cite{Nielsen}.
\begin{definition}[\textbf{The purity of a quantum state}]
\label{eqn:purity}
The map $\gamma(\boldsymbol{\hat{\rho}}):= \mathrm{Tr}\big\{\boldsymbol{\hat{\rho}^{2}}
 \big\}$ is known as the \emph{purity} \cite{Nielsen}, and it is one of many measures of mixedness for density operators. For any density operator $\boldsymbol{\hat{\rho}}$ acting in a Hilbert space $\mathscr{H}$ (possibly infinite-dimensional), the purity is bounded as follows.   
\begin{equation}
 \frac{1}{\mathbf{dim}\big\{\mathscr{H}\big\}}\leq\gamma\big( \boldsymbol{\hat{\rho}}\big)\leq 1. 
\end{equation}
\end{definition}  
Now, the equation 
\begin{equation}
\label{eqn:liouville}
\partial_{t}\boldsymbol{\hat{\rho}}_{t}=-i\big[\mathbf{\hat{H}},\: \boldsymbol{\hat{\rho}}_{t}\big] 
\end{equation}
where  $\boldsymbol{\hat{\rho}}_{0} \in \mathsf{S}(\mathscr{H})$ is given, is known as the Liouville-von Neumann (LvN) equation  \cite{han}. The LvN equation is of great importance to quantum theory as it is the operator version of the Schr\"{o}dinger equation. The solution to the LvN equation $\boldsymbol{\hat{\rho}}_{t}$ is indeed a density operator \cite{Townsend} \cite{Nielsen}. In the following section we will further formalize the LvN equation, deducing the appropriate domain of the solutions $\boldsymbol{\hat{\rho}}_{t}$.
\end{section}

\begin{section}{Ergodic averages and more on semigroups}
\label{eqn:sec3}
\;\;\; Given a strongly continuous one-parameter semigroup (Definition \ref{eqn:semi}) $\mathbf{\hat{T}}_{t}$ \cite{yos} \cite{lax}, we define the ergodic average of said semigroup with respect to the parameter $t$ as follows.

\begin{definition}[\textbf{Cesaro mean, aka} $T$-\textbf{time average} ]
\label{eqn:terg}
Let be $\mathbf{\hat{T}}_{t}$ a $c_{0}$-semigroup acting on a Banach space $\mathscr{B}$. We define the Cesaro means of the semigroup $\mathbf{\hat{T}}_{t}$ at time $T>0$ as follows. 
\begin{equation}
    \big\langle\!\!\big\langle\mathbf{\hat{T}}_{t}\big\rangle\!\!\big\rangle_{T}:= \frac{1}{T}\int_{0}^{T}\mathbf{\hat{T}}_{t}dt,
    \end{equation}
where the operators are defined in the sense of Bochner integrals (Definition \ref{eqn:boch}).  
\end{definition}

From the above definition, we can define the ergodic average of the semigroup $\mathbf{\hat{T}}_{t}$ as follows.

\begin{definition}[\textbf{Ergodic average}]
Consider a $c_{0}$- semigroup $\mathbf{\hat{T}}_{t}$ acting on a Banach space $\mathscr{B}$. We define the ergodic average of $\mathbf{\hat{T}}_{t}$ with respect to $t$ as follows. 
\begin{equation}
\big\langle\!\!\big\langle\mathbf{\hat{T}}_{t}\big\rangle\!\!\big\rangle_{\infty}:= \lim_{T\rightarrow \infty} \big\langle\!\!\big\langle\mathbf{\hat{T}}_{t}\big\rangle\!\!\big\rangle_{T}
\end{equation}
where the limit is taken with respect to the strong operator norm topology (Definition \ref{eqn:sot}.) 
\end{definition}
Necessary and sufficient conditions for the ergodic average to exist are known; the reader is referred to \cite{kiesenhofer} for a compact synopsis. 
When $\lim_{T\rightarrow \infty} \langle\!\langle\mathbf{\hat{T}}_{t}\rangle\!\rangle_{T}$ exists, the semigroup $\mathbf{\hat{T}}_{t}$ is said to be \textbf{Mean Ergodic}. The following theorems showcase what the afforementioned necessary and sufficient conditions for mean ergodicity are \cite{kiesenhofer}.
\begin{definition4}
\label{eqn:theoremc0}
A $c_{0}$-semigroup $\mathbf{\hat{T}}_{t}$ on a Banach space $\mathscr{B}$ satisfying 
\begin{equation}
    \lim_{t\rightarrow \infty}\frac{\|\mathbf{\hat{T}}_{t}\|}{t} = 0
\end{equation}
is mean ergodic if and only if  
\begin{equation}
    \mathscr{B} = \mathrm{Fix}\big(\mathbf{\hat{T}}_{t}\big) \bigoplus \overline{\big\{\mathbf{\hat{v}}-\mathbf{\hat{T}}_{t}\mathbf{\hat{v}} \big| \mathbf{\hat{v}}\in \mathscr{B}, t\geq 0\big\}}
\end{equation}
where $\mathrm{Fix}$ refers to the fixed points of the semigroup $\mathbf{\hat{T}}_{t}$. 
\end{definition4}
The latter leads to the following corollary.
\begin{Co}{A theorem from \cite{kiesenhofer}}
\label{eqn:corollc0}
A $c_{0}$-semigroup $\mathbf{\hat{T}}_{t}$ on a Banach space $\mathscr{B}$, with generator $\mathbf{\hat{H}}$ satisfying 
\begin{equation}
    \lim_{t\rightarrow \infty}\frac{\big\|\mathbf{\hat{T}}_{t}\big\|}{t} = 0
\end{equation}
is mean ergodic if and only if  the fixed space 
\begin{equation}
    \mathrm{Fix}\big( \mathbf{\hat{T}}_{t} \big) = \mathrm{Ker}\big(\mathbf{\hat{H}}\big) 
\end{equation}
separates the dual fixed space 
\begin{equation}
    \mathrm{Fix}\big( \mathbf{\hat{T}}_{t}^{'} \big) = \mathrm{Ker}\big(\mathbf{\hat{H}}^{'}\big) 
\end{equation}
where the notation $\mathbf{\hat{A}}^{'}$ is used to symbolize the dual operator corresponding to $\mathbf{\hat{A}}$.
\end{Co}
There is also a weaker sufficient condition condition that may be applied for the case of weakly-compact semigroups \cite{kiesenhofer}. 

\begin{definition4}{A theorem from \cite{kiesenhofer}}
\label{eqn:theorem2c0}
A $c_{0}$-semigroup $\mathbf{\hat{T}}_{t}$ on a Banach space $\mathscr{B}$ satisfying 
\begin{equation}
    \lim_{t\rightarrow \infty}\frac{\big\|\mathbf{\hat{T}}_{t}\big\|}{t} = 0
\end{equation}
is mean ergodic if and only if  
for all $\mathbf{\hat{v}}\in \mathscr{B}$ there exists a sequence $t_{n}\rightarrow \infty$ such that $ \langle\!\langle \mathbf{\hat{T}}_{t} \rangle\!\rangle_{t_{n}}\mathbf{\hat{v}}$ converges in the weak topology of $\mathscr{B}$ (see \cite{Simon} for a definition of weak topology).
\end{definition4}
In the case of quantum dynamics, the semigroups used are those generated by self-adjoint operator $\mathbf{\hat{H}}$, namely $\mathbf{\hat{U}}_{t}:=e^{-it\mathbf{\hat{H}}}$ \cite{gerald}. Regardless of the spectral properties of the operator $\mathbf{\hat{H}}$, it is easily verified that the growth-condition $\lim_{t\rightarrow \infty}t^{-1}\|\mathbf{\hat{T}}_{t}\| = 0$ immediately follows as a consequence of the unital norm of unitary operators, i.e. $\|\mathbf{\hat{U}}_{t}\| =1$. Assuming that we know the spectral decomposition of the generator $\mathbf{\hat{H}}$, the Hamiltonian, the if and only if conditions from Theorem \ref{eqn:theoremc0}, Theorem \ref{eqn:theorem2c0} and  Corollary \ref{eqn:corollc0} may also be easily verified. The latter results do not, however, tell us what is the limit of $\langle\!\langle\mathbf{\hat{T}}_{t}\rangle\!\rangle_{T}$ converges to as $T\rightarrow \infty$. To deduce this limit in a concrete manner within the framework corresponding to quantum theory and unitary continuous semigroups, we must apply a result specialized to Hilbert spaces. Namely, the so-called  von Neaumann ergodic theorem \cite{Simon}. Prior to presenting this theorem, note that for discrete-time the ergodic average in question is 
\begin{equation}
\label{eqn:head1}
\lim_{n\rightarrow \infty}\frac{1}{n}\sum_{k=0}^{n-1}\mathbf{\hat{U}}_{k}
\end{equation}
This is the ergodic average most often encountered in the literature on ergodic theory \cite{Simon}. The continuous time ergodic average is of course 
\begin{equation}
\label{eqn:head2}
 \lim_{T\rightarrow \infty} \big\langle\!\!\big\langle\mathbf{\hat{U}}_{t}\big\rangle\!\!\big\rangle_{T}
\end{equation}
where we taken the limit with respect to the strong operator norm topology. We now present the von Neumann theorem, a theorem whose power lies in its ability to determine such limits as (\ref{eqn:head1}) and (\ref{eqn:head2})when they exist \cite{Simon}. 

\begin{definition4}[\textbf{von Neumann's Ergodic Theorem}]
\label{eqn:vonerg}
Let $\mathbf{\hat{U}}_{t}$ be a unitary operator on a Hilbert space $\mathscr{H}$. Let $\mathbf{\hat{P}}$ be the orthogonal projection onto $\mathrm{Ker}\big(\mathbf{\hat{I}}-\mathbf{\hat{U}}_{1}\big)=\big\{ |\psi\rangle \in \mathscr{H} \big| \mathbf{\hat{U}}_{1}|\psi\rangle = |\psi\rangle\big\}$. Then, for any $|\psi\rangle\in\mathscr{H}$, we have:
\begin{equation}
\lim_{T\rightarrow \infty} \big\langle\!\!\big\langle\mathbf{\hat{U}}_{t}\big\rangle\!\!\big\rangle_{T}| \psi\rangle= \mathbf{\hat{P}} |\psi\rangle
\end{equation}
where the limit is taken with respect to the norm on $\mathscr{H}$. In other words, the T-time averages $\langle\!\langle\mathbf{\hat{U}}_{t}\rangle\!\rangle_{T}$ converge to $\mathbf{\hat{P}}$ with respect to the strong operator topology induced by the operator norm. The same result follows for the finite time ergodic mixture (\ref{eqn:head1}).
\end{definition4}

In von Neumann's seminal work on ergodic theory, namely \cite{vonneu1} \cite{vonneu2}, he actually studied the convergence of the operators 
\begin{equation}
\frac{1}{T-S}\int_{S}^{T}\mathbf{\hat{U}}_{t}dt
\end{equation}
as $T-S \rightarrow \infty$ irrespectively of he mode of variation of $S, T$. In this sense, von Neumann's result was a bit more general. Now,  recalling that Theorem \ref{eqn:stone} establishes a one-t-one correspondence between unitary $c_{0}$-semigroups and self-adjoint operators, it is not very restrictive to consider an example where $\mathscr{H}=\mathbb{C}^{d}$ and the unitary semigroup $\{\mathbf{\hat{U}}_{t}\}_{t\in\mathbb{R}_{+}}$ has a non-degenerate self-adjoint generator $\mathbf{\hat{H}}$ with spectral decomposition 
\begin{equation}
    \mathbf{\hat{H}} = \sum_{i=1}^{d} \lambda_{i}\mathbf{\hat{P}}_{i}
\end{equation}
We can further consider two cases, one where $\mathbf{\hat{H}}$ has no eigenvectors with eigenvalue zero and one in which it does. It is clear from Theorem \ref{eqn:vonerg} that $\mathbf{\hat{U}}_{1}|\psi\rangle = |\psi\rangle$ only when $|\psi\rangle \in \mathrm{Ker}(\mathbf{\hat{H}})$, whence $\mathrm{Ker}(\mathbf{\hat{H}}) = \mathrm{Ker}(\mathbf{\hat{I}}-\mathbf{\hat{U}}_{1}) = \{\emptyset \}$. It therefore follows that 
\begin{equation}
\lim_{T\rightarrow \infty}\big\langle\!\!\big\langle\mathbf{\hat{U}}_{t}\big\rangle\!\!\big\rangle_{T} = 0
\end{equation}
when $\mathrm{Ker}(\mathbf{\hat{H}})$ is empty, otherwise the limit is simply a projection onto the kernel of the generator $\mathbf{\hat{H}}$,
i.e. in the limit only the fixed points survive while everything else projects to zero. One might be tempted to interpret the limits 
\begin{equation}
\lim_{T\rightarrow \infty} \big\langle\!\!\big\langle\mathbf{\hat{U}}_{t}\big\rangle\!\!\big\rangle_{T}\big|\psi\big\rangle
\end{equation}
as the average state of some system given that $T$ time has elapsed, as $T$ approaches infinity. Nevertheless, the vector 
\begin{equation}
\lim_{T\rightarrow \infty} \big\langle\!\!\big\langle\mathbf{\hat{U}}_{t}\big\rangle\!\!\big\rangle_{T}\big|\psi\big\rangle
\end{equation}
is not normalized and hence cannot be a quantum state. However, the latter may be viewed as a type of non-unitary dynamics where information or energy is lost to some environment.

\;\;\;
 To illustrate how one may apply the von Neumann ergodic theorem to the case pertaining to a density-operator-description of quantum mechanics let $dim(\mathscr{H})=d<\infty$ and let the Hamiltonian $\mathbf{\hat{H}}$ have non-degenerate pure-point spectrum \cite{Simon} \cite{lax} \cite{gerald}; expressed in spectral form $\mathbf{\hat{H}} = \sum_{n=1}^{d}\lambda_{n}\mathbf{\hat{P}}$. The abstract solution of the LvN equation (\ref{eqn:liouville}) in this case is $\mathbf{\hat{U}}_{t}\boldsymbol{\hat{\rho}}_{0}\mathbf{\hat{U}}_{t}^{\dagger}$. Here, $\boldsymbol{\hat{\rho}}_{0}$ is the initial state of the system and $\{\mathbf{\hat{U}}_{t}\}_{t\in\mathbb{R}_{+}}$is the unitary $c_{0}$-semigroup generated by $\mathbf{\hat{H}}$. In such a case the respective ergodic averages are of the following form.
\begin{equation}
\label{eqn:erglim}
\big\langle\!\!\big\langle\boldsymbol{\hat{\rho}}_{t} \big\rangle\!\!\big\rangle_{\infty}:=\lim_{T\rightarrow \infty} \big\langle\!\!\big\langle\mathbf{\hat{U}}_{t}\boldsymbol{\hat{\rho}}_{0}\mathbf{\hat{U}}_{t}^{\dagger}\big\rangle\!\!\big\rangle_{T}; \;\big(\boldsymbol{\hat{\rho}}_{t}:=\mathbf{\hat{U}}_{t}\boldsymbol{\hat{\rho}}_{0}\mathbf{\hat{U}}_{t}^{\dagger}\big)
\end{equation}
It is easy to see that the resulting object is also a density operator. This is done by noting that the trace is preserved; i.e. 
\begin{equation}
\mathrm{Tr}\left\{\lim_{T\rightarrow \infty} \big\langle\!\!\big\langle\mathbf{\hat{U}}_{t}\boldsymbol{\hat{\rho}}_{0}\mathbf{\hat{U}}_{t}^{\dagger}\big\rangle\!\!\big\rangle_{T} \right\} = \lim_{T\rightarrow \infty}\frac{1}{T}\int_{0}^{T}\mathrm{Tr}\{\mathbf{\hat{U}}_{t}\boldsymbol{\hat{\rho}}_{0}\mathbf{\hat{U}}_{t}^{\dagger}\}dt = 1,
\end{equation}
owing to the trace being preserved under unitary transformations.
Positive definiteness can be proven by noting that the map $\boldsymbol{\hat{\rho}}_{t} \to \langle\!\langle\mathbf{\hat{U}}\boldsymbol{\hat{\rho}}_{t} \mathbf{\hat{U}}^{\dagger} \rangle\!\rangle_{T}$ is a completely positive map \cite{Nielsen}. Furthermore, by applying Theorem \ref{eqn:vonerg}, the limit (\ref{eqn:erglim}) can be determined. To that end, let us rewrite the solution to the LvN equation $\mathbf{\hat{U}}_{t}\boldsymbol{\hat{\rho}}_{0}\mathbf{\hat{U}}_{t}^{\dagger}$.To see the equivalence between these two expression notice that the LvN equation is just a case of the abstract Cauchy problem \cite{han}
\begin{equation}
\label{eqn:liou}
\partial_{t}\boldsymbol{\hat{\rho}}_{t} = \mathfrak{\hat{O}}_{\mathbf{\hat{H}}}\boldsymbol{\hat{\rho}}_{t}
\end{equation}
where the "super-operator" $\mathfrak{\hat{O}}_{\mathbf{\hat{H}}}\boldsymbol{\hat{\rho}}:=-i[\mathbf{\hat{H}},\boldsymbol{\hat{\rho}}]$ can be viewed as the generator of the semigroup associated with the respective abstract Cauchy-problem (\ref{eqn:liou}), whence, $e^{-t\mathfrak{\hat{O}}_{\mathbf{\hat{H}}}}$ is formally the dynamical semigroup which evolves an initial state $\boldsymbol{\hat{\rho}}_{0}$ at $t= 0 $ to $\boldsymbol{\hat{\rho}}_{t}$. In general, i.e. for $\mathscr{H}$ of arbitrary dimension, for the semigroup  $e^{-t\mathfrak{\hat{O}}_{\mathbf{\hat{H}}}}$ and the Cauchy-Problem (\ref{eqn:liou}) to be well-defined, one must carefully choose the domain of the superoperator $\mathfrak{\hat{O}}_{\mathbf{\hat{H}}}$ to be densely defined in an appropriate Hilbert-Space. The super-operator in question is then the closure of said densely defined operator \cite{han}. For the case of the LvN equation, it is natural to select the space of Hilbert-Schmidt operators as the space of slutions; this is because each $\boldsymbol{\hat{\rho}}$ describing the sate of a quantum system may be identified with the tensor-product space $\mathcal{L}(\mathscr{H}):=\mathscr{H}\bigotimes \mathscr{H}^{*}$, referred to as Liouville space in this context \cite{han}, where $\mathscr{H}$ is a Hilbert space and $\mathscr{H}^{*}$ of course its dual. Indeed, $\mathcal{L}(\mathscr{H})$ equipped with the Hilbert-Schmidt inner product $\langle \mathbf{\hat{A}}, \mathbf{\hat{B}}\rangle_{H.S.}:=Tr\{\mathbf{\hat{A}}^{\dagger}\mathbf{\hat{B}}\}$ constitutes a Hilbert space. The Hilbert-Schmidt inner product induces a metric called the Hilbert-Schmidt distance. Let $\boldsymbol{\hat{\rho}}$ and $\boldsymbol{\hat{\sigma}}$ be two density operators, the Hilbert-Schmidt distance is defined as follows.
\begin{equation}
\big\|\boldsymbol{\hat{\rho}}-\boldsymbol{\hat{\sigma}}\big\|_{H.S} := \sqrt{\big\langle \boldsymbol{\hat{\rho}}-\boldsymbol{\hat{\sigma}}, \boldsymbol{\hat{\rho}}-\boldsymbol{\hat{\sigma}}\big\rangle_{H.S.}} 
\end{equation} 
With the latter in play, we may now make sense of the limit (\ref{eqn:erglim}) for the semigroup $e^{-t\mathfrak{\hat{O}}_{\mathbf{\hat{H}}}} = \mathbf{\hat{U}}_{t}(\cdot)\mathbf{\hat{U}}_{t}^{\dagger}$. Continuing with our example, we remind the reader that we assume that for now we assume that $dim(\mathscr{H})<\infty$.
Now, define 
\begin{equation}
\Lambda: = \big\{\mathbf{\hat{A}} \in \mathscr{S}_{2}\big(\mathscr{H}\big) | e^{-t\mathfrak{\hat{O}}_{\mathbf{\hat{H}}}}\mathbf{\hat{A}} = \mathbf{\hat{A}}\big\} = \mathrm{Ker}\big(\mathbf{\hat{I}}- e^{-t\mathfrak{\hat{O}}_{\mathbf{\hat{H}}}} \big)
\end{equation}
then, by Theorem \ref{eqn:vonerg},
\begin{equation}
    \lim_{T\rightarrow \infty} \big\langle\!\!\big\langle e^{-t\mathfrak{\hat{O}}_{\mathbf{\hat{H}}}} \boldsymbol{\hat{\rho}}_{0}\big\rangle\!\!\big\rangle_{T} = \mathbf{\hat{P}}_{\Lambda}\boldsymbol{\hat{\rho}}_{0}
\end{equation}
for any state $\boldsymbol{\hat{\rho}}_{0}\in\mathscr{S}_2(\mathscr{H})$, where the limit is taken with respect to the Hilbert-Schmidt norm $\|\cdot\|_{HS}$ and $\mathbf{\hat{P}}_{\Lambda}$ is the projector onto the subspace $\Lambda$. Here, 
\begin{equation}
    \lim_{T\rightarrow \infty} \big\langle\!\!\big\langle e^{-t\mathfrak{\hat{O}}_{\mathbf{\hat{H}}}} \big\rangle\!\!\big\rangle_{T} = \mathbf{\hat{P}}_{\Lambda}
 \end{equation}
with respect to the strong operator topology generated on $L(\mathscr{S}_2(\mathscr{H})),$ the set of bounded linear maps on the Hilbert space $\mathscr{S}_2(\mathscr{H}),$ by the Hilbert-Schmidt norm (\ref{eqn:hilsch}). Furthermore, it is evident that for $e^{-t\mathfrak{\hat{O}}_{\mathbf{\hat{H}}}} \mathbf{\hat{A}} = \mathbf{\hat{A}}$ to hold, $[\mathbf{\hat{H}},\mathbf{\hat{A}}] = 0$ must be satisfied. In other words, given that $\mathbf{\hat{H}} = \sum_{n=1}^{d}\lambda_{n}\mathbf{\hat{P}}_{n}$, we require that $\sum_{n=1}^{d}\lambda_{n}\big[\mathbf{\hat{P}}_{n}, \mathbf{\hat{A}}\big] = 0 $, which implies that $\big[\mathbf{\hat{P}}_{n}, \mathbf{\hat{A}}\big] = 0$ for all $n\in[1,...,d]$ due to the linear independence of the projectors $\mathbf{\hat{P}}_{n}$. However, the only way that this may be so is if $\mathbf{\hat{A}} = \sum\mu_{n}\mathbf{\hat{P}}_{n}$, i.e. $\mathbf{\hat{A}}$ and $\mathbf{\hat{H}}$ are simultaneously diagonalizable. This leads us to conclude the ensuing claim.
\begin{definition6}
For the finite-dimensional setting developed above 
\begin{equation}
\lim_{T\rightarrow \infty}\big\langle\!\!\big\langle e^{-t\mathfrak{\hat{O}}_{\mathbf{\hat{H}}}} \boldsymbol{\hat{\rho}}_{0}\big\rangle\!\!\big\rangle_{T} = \sum_{n=1}^{d}\mathbf{\hat{P}}_{n}\boldsymbol{\hat{\rho}}_{0}\mathbf{\hat{P}}_{n}
\end{equation}
\end{definition6}
\begin{proof}
Let us begin with the definition of $\mathbf{\hat{P}}_{\Lambda}$. \begin{equation}
\mathbf{\hat{P}}_{\Lambda}: = \big\{\mathbf{\hat{A}} \in \mathscr{S}_{2}\big(\mathscr{H}\big)\big| e^{-t\mathfrak{\hat{O}}_{\mathbf{\hat{H}}}}\mathbf{\hat{A}} = \mathbf{\hat{A}}\big\}
\end{equation}
Indeed, for finite-dimensional $\mathscr{H}$ and full-rank non-degenerate $\mathbf{\hat{H}}$, $\mathbf{\hat{A}}\in \Lambda$ must belong to the direct sum Hilbert space 
\begin{equation}
\bigoplus_{n}\mathscr{H}_{n}\otimes\mathscr{H}_{n}^{*}
\end{equation}
where each of the $\mathscr{H}_{n}$ is a one-dimensional subspace of $\mathscr{H}$ invariant under $\mathbf{\hat{H}}$. As such, the operator $\mathbf{\hat{A}}$ may be uniquely decomposed via orthogonal projections afforded by the Hilbert-Schmidt inner product. Let $\mathbf{\hat{p}}_{n}$ be a projector onto the subspace $\mathscr{H}_{n}$ and $\mathbf{\hat{p}}_{n}^{*}$ a projector onto the dual space $\mathscr{H}_{n}^{*}$. The projector onto the subspace $\mathscr{H}_{n}\otimes\mathscr{H}_{n}^{*}$ is therefore
\begin{equation}
\mathbf{\hat{P}}_{\mathscr{H}_{n}\otimes\mathscr{H}_{n}^{*}}\mathbf{\hat{A}}=\mathbf{\hat{p}}_{n}\otimes\mathbf{\hat{p}}_{n}^{*}\big\langle \mathbf{\hat{p}}_{n}\otimes\mathbf{\hat{p}}_{n}^{*}, \mathbf{\hat{A}}\big\rangle_{H.S.} = \Big(\mathbf{\hat{p}}_{n}\otimes\mathbf{\hat{p}}_{n}^{*}\Big)^{'}\mathbf{\hat{A}}\Big(\mathbf{\hat{p}}_{n}\otimes\mathbf{\hat{p}}_{n}^{*}\Big)
\end{equation}
therefore
\begin{equation}
    \mathbf{\hat{A}} = \sum_{n}\lambda_{n}\mathbf{\hat{p}}_{n}\otimes\mathbf{\hat{p}}_{n}^{*} 
\end{equation}
where $\lambda_{n}:= \big\langle \mathbf{\hat{p}}_{n}\otimes\mathbf{\hat{p}}_{n}^{*}, \mathbf{\hat{A}}\big\rangle_{H.S.}=\Big(\mathbf{\hat{p}}_{n}\otimes\mathbf{\hat{p}}_{n}^{*}\Big)^{'}$; here the prime $'$ symbolizes the dual of a bounded operator. Defining $\mathbf{\hat{P}}_{n}:=\mathbf{\hat{p}}_{n}\otimes\mathbf{\hat{p}}_{n}^{*}$, corresponding to the eigensubspaces of $\mathbf{\hat{H}}$, we have 
\begin{equation}
    \mathbf{\hat{A}} = \sum_{n}\lambda_{n}\mathbf{\hat{P}}_{n}
\end{equation}
which may be rewritten as 
\begin{equation}
    \mathbf{\hat{A}} = \sum_{n}\mathbf{\hat{P}}_{n}^{'}\mathbf{\hat{A}}\mathbf{\hat{P}}_{n}
\end{equation}
\;\;\;In physics notation, where $\mathbf{\hat{A}}$ has the decomposition
$\mathbf{\hat{A}} = \sum_{n,m}\alpha_{n,m}\big|\psi_{n}\big\rangle\big\langle\psi_{m}\big|$
,using the Dirac notation, one may simply write 
\begin{equation}
\label{eqn:projproj}
    \mathbf{\hat{A}} = \sum_{n}\mathbf{\hat{P}}_{n}\mathbf{\hat{A}}\mathbf{\hat{P}}_{n} = \sum_{n}\big|\phi_{n}\big\rangle\big\langle \phi_{n}\big|\mathbf{\hat{A}}\big|\phi_{n}\big\rangle\big\langle \phi_{n}\big| = \sum_{n}\big\langle \phi_{n}\big|\mathbf{\hat{A}}\big|\phi_{n}\big\rangle\big|\phi_{n}\big\rangle\big\langle \phi_{n}\big|
\end{equation}
where we used the definition $\mathbf{\hat{P}}_{n}:= |\phi_{n}\rangle\langle \phi_{n}|$, making it unnecessary for us to be explicit about the fact that projectors acting on the right of $\mathbf{\hat{A}}$ in (\ref{eqn:projproj}) are dual operators of the $\mathbf{\hat{P}}_{n}$.
\end{proof}

\end{section}
\begin{section}{Equilibration-on-Average}
\label{eqn:sec4}
\;\;\; As discussed in the previous section, when studying how a quantum mechanical system equilibrates from a unitary-dynamics perspective, one of the central items of study is the following completely positive trace preserving map \cite{re1} \cite{short} \cite{re2}.   
\begin{equation}
\label{eqn:marcelo}
\lim_{T\rightarrow\infty}\big\langle\!\!\big\langle e^{-t\mathfrak{\hat{O}}_{\mathbf{\hat{H}}}}\big(\cdot\big)\big\rangle\!\!\big\rangle_{T}
\end{equation}
When such a limit exists, and the limiting super operator takes any initial quantum state $\boldsymbol{\hat{\rho}}_{0}$ to one that is invariant under the quantum map (\ref{eqn:marcelo}), one can use such state to study the \emph{equilibration-on-average} of observables with respect to the dynamics generated by the semigroup $e^{-t\mathfrak{\hat{O}}_{\mathbf{\hat{H}}}}$. The following definition formalizes these notions \cite{short}. 
\begin{definition}[Equilibration-on-Average of an observable]
Consider a system evolving unitarily that is described by the density operator $\boldsymbol{\hat{\rho}}_{t} \in \mathcal{S}(\mathscr{H})$ and let $\mathbf{\hat{A}}\in L(\mathscr{H})$. $\mathbf{\hat{A}}$ is said to \emph{equlibrate-on-average} with respect to the dynamical quantum state $\boldsymbol{\hat{\rho}}_{t}$ if
\begin{equation}
\label{eqn:noper}
\sigma_{\mathbf{\hat{A}}}^{2}(\infty):=\lim_{T\rightarrow \infty}\Big\langle\!\!\!\Big\langle\Big(Tr\big\{\mathbf{\hat{A}}\boldsymbol{\hat{\rho}}_{t}\big\}- Tr\big\{\mathbf{\hat{A}}\big\langle\!\!\big\langle\boldsymbol{\hat{\rho}}_{t}\big\rangle\!\!\big\rangle_{\infty}\big\}\Big)^{2}\Big\rangle\!\!\!\Big\rangle_{T} = 0 
\end{equation}
\end{definition}
The term $\sigma_{\mathbf{\hat{A}}}^{2}(\infty)$ may be interpreted as follows. An observable $\mathbf{\hat{A}}$ equilibrates on average with respect to the state $\boldsymbol{\hat{\rho}}_{t}$, when, for most times t, a measurement of the observable will follow statistics close to those of the equilibrium state $\langle\!\langle\boldsymbol{\hat{\rho}}_{t}\rangle\!\rangle_{\infty}$. Of course, achieving perfect equilibration-on-average will generally not be feasible, hence estimates of the long-time average error $\sigma_{\mathbf{\hat{A}}}^{2}(\infty)$ will be necessary to understand how small this quantity may be; the smallness is most importantly characterized by the so-called effective dimension $d_{eff}$ defined in the following theorem. Perhaps the most well-known is the following result by Short \cite{short}. 
\begin{definition4}[\textbf{Result on equilibration estimates by A.J. Short}\cite{short}]
\label{eqn:themainresultriemann}
Consider a d-dimensional quantum system evolving under a Hamiltonian $\mathbf{\hat{H}} = \sum_{n}\lambda_{n}\mathbf{\hat{P}}_{n}$, where $\mathbf{\hat{P}}_{n}$ is the projector onto the eigenspace with eigenvalue $\lambda_{n}$. Denote the system's density operator $\boldsymbol{\hat{\rho}}_{t}$, and its ergodic average by $\langle\!\langle \boldsymbol{\hat{\rho}}_{t}\rangle\!\rangle_{\infty}$. If $\mathbf{\hat{H}}$ has non-degenerate energy gaps, then for any operator $\mathbf{\hat{A}}$
\begin{equation}
0\leq\Big\langle\!\!\!\Big\langle\Big(Tr\big\{\mathbf{\hat{A}}\boldsymbol{\hat{\rho}}_{t}\big\}- Tr\big\{\mathbf{\hat{A}}\big\langle\!\!\big\langle\boldsymbol{\hat{\rho}}_{t}\big\rangle\!\!\big\rangle_{\infty}\big\}\Big)^{2}\Big\rangle\!\!\!\Big\rangle_{\infty}\leq \frac{\Delta\big(\mathbf{\hat{A}}\big){2}}{4d_{eff}}\leq \frac{\big\|\mathbf{\hat{A}}\big\|^{2}}{d_{eff}}
\end{equation}
where 
\begin{equation}
d_{eff}:= \frac{1}{\sum_{n}\big(Tr\big\{\mathbf{\hat{P}}_{n}\boldsymbol{\hat{\rho}}_{0}\big\}\big)^{2}} 
\end{equation}
and
\begin{equation}
    \Delta\big(\mathbf{\hat{A}}\big):= 2 \min_{c\in\mathbb{C}}\big\|\mathbf{\hat{A}}-c\mathbf{\hat{I}}\big\|
\end{equation}
\end{definition4}
\end{section}
It is worth highlighting that the effective dimension $d_{eff}$ depends on the spectral properties of $\boldsymbol{\hat{\rho}}_{0}$ and in particular on the relative purity (Definition \ref{eqn:purity}) of the output of the corresponding fully-dephasing map $\sum_{n}\mathbf{\hat{P}}_{n}\boldsymbol{\hat{\rho}}_{0}\mathbf{\hat{P}}_{n}$. This can be seen by noting that $\gamma( \sum_{n}\mathbf{\hat{P}}_{n}\boldsymbol{\hat{\rho}}_{0}\mathbf{\hat{P}}_{n})=\sum_{n}(Tr\{\mathbf{\hat{P}}_{n}\boldsymbol{\hat{\rho}}_{0}\})^{2}$. As an example of $d_{eff}$ consider the case where $\boldsymbol{\hat{\rho}}_{0} = \sum_{n=1}^{d}\frac{1}{d}\mathbf{\hat{P}}_{n}$. In this case, $d_{eff} =d$, suggesting that the possibility of exact equilibration-on-average becomes more likely as the dimension of the Hilbert space $\mathscr{H}$ becomes larger. Returning to the setting of Theorem \ref{eqn:themainresultriemann}, the likeliness of $d_{eff}$ being large in this case will of course depend on the relationship between the $\mathbf{\hat{P}}_{n}$ and the initial state $\boldsymbol{\hat{\rho}}_{0}$. To the extent that the authors are informed, up until the moment of the submission of this work, there have been no attempts to develop a version of Theorem \ref{eqn:themainresultriemann} for the case where the Hamiltonian $\mathbf{\hat{H}}$ acts on an infinite-dimensional Hilbert space; a case which introduces the possibility of $\mathbf{\hat{H}}$ having spectra which do not constitute a point-spectrum (the only kind of spectrum in the finite-dimensional case), such as continuous spectra \cite{Simon} \cite{lax} \cite{gerald}. We will explore these notions in the following section, focusing our attention on the case where $\mathbf{\hat{H}}$ has purely continuous-spectrum.

\newpage 

\begin{section}{A first attempt at studying equilibration for the case of continuous variables}
\label{eqn:sec5}
\;\;\; Let us now consider a setting analogous to the one depicted in Theorem \ref{eqn:themainresultriemann}, but now with a self-adjoint operator $\mathbf{\hat{H}}$ having purely continuous spectrum (see Definition \ref{eqn:thespectrum}) since this is where the uncharted territory lies. The focus of this section will be to obtain a result analogous to (\ref{eqn:themainresultriemann}) for such a case. The case where the Hilbert space $\mathscr{H}$ in question is infinite dimensional and the Hamiltonian $\mathbf{\hat{H}}$ has purely pure-point spectrum may be studied with Theorem \ref{eqn:themainresultriemann}; the only modification that one would have to make is to restrict the set of possible observables to that of bounded self-adjoint operators. Firstly, we must point out that when $\mathbf{\hat{H}}$ has purely continuous spectrum, the limit $\lim_{T\rightarrow\infty}\langle\!\langle e^{-t\mathfrak{\hat{O}}_{\mathbf{\hat{H}}}}(\cdot)\rangle\!\rangle_{T}$
converges to zero in the strong operator topology induced by the Hilbert-Schmidt norm. This follows immediately from the von Neumann Ergodic theorem (Theorem \ref{eqn:vonerg}). In this case, $\lim_{T\rightarrow\infty}\langle\!\langle e^{-t\mathfrak{\hat{O}}_{\mathbf{\hat{H}}}}(\cdot)\rangle\!\rangle_{T}$ converges to a projector which takes any density operators $\boldsymbol{\hat{\rho}}_{0}$ and projects it onto the subspace of the corresponding Liouville space $\mathcal{L}(\mathscr{H})$ invariant under the dynamics pertaining to the semigroup $e^{-t\mathfrak{\hat{O}}_{\mathbf{\hat{H}}}}$; i.e. $\boldsymbol{\hat{\rho}}_{0}$ such that $e^{-t\mathfrak{\hat{O}}_{\mathbf{\hat{H}}}}\boldsymbol{\hat{\rho}}_{0} = \boldsymbol{\hat{\rho}}_{0}$. However, the latter is so if and only if $ [\mathbf{\hat{H}},\boldsymbol{\hat{\rho}}_{0}] = 0  $, which is impossible since $\boldsymbol{\hat{\rho}}_{0}$ is compact while $\mathbf{\hat{H}}$ is not \cite{Simon} \cite{lax}, meaning that these two operators cannot be written in spectral form (\ref{eqn:specform}) with respect to the same spectral measure; this should come as no surprise since $\mathbf{\hat{H}}$ has purely continuous spectrum while density operators $\boldsymbol{\hat{\rho}}$ have purely point spectrum. This means that taking the limit $T\rightarrow \infty$ is not physical, aka not stable, for the case where $\mathbf{\hat{H}}$ has purely continuous spectrum. We will therefore replace the map $\langle\!\langle\cdot\rangle\!\rangle_{\infty}$ with one assuming a finite final time $T$, the $T-$time average, aka $T$-time ergodic average (Definition \ref{eqn:terg}), aka Cesaro mean, and study this quantum map instead.

\subsection{Spectral families of projection operators}

\;\;\;The spectral theorem says that a self-adjoint operator $\mathbf{\hat{H}}$ may be written in the following so-called spectral form \cite{jordan}.
\begin{equation}
\label{eqn:specform}
\mathbf{\hat{H}} = \int_{\sigma(\mathbf{\hat{H}})}\lambda\mathbf{\hat{E}}(d\lambda)
\end{equation}
Here the operators $\mathbf{\hat{E}}(\lambda)$ constitute the unique spectral family \cite{jordan} of projectors associated with the operator $\mathbf{\hat{H}}$.
\begin{definition}[\textbf{Spectral families of projection operators}]
A family of projection operators $\mathbf{\hat{E}}(\lambda)$ depending on a real parameter $\lambda$ is a spectral family if it has the following properties:
\begin{itemize}
\item If $\lambda\leq \mu$, then $\mathbf{\hat{E}}(\lambda)\leq \mathbf{\hat{E}}(\mu)$ or $\mathbf{\hat{E}}(\lambda)\mathbf{\hat{E}}(\mu)=\mathbf{\hat{E}}(\lambda) = \mathbf{\hat{E}}(\mu)\mathbf{\hat{E}}(\lambda)$

\item If $\varepsilon$ is positive, $\mathbf{\hat{E}}(\lambda+\varepsilon)\big|\psi\big\rangle \rightarrow \mathbf{\hat{E}}(\lambda)\big|\psi\big\rangle $ as $\varepsilon\rightarrow 0$ for any vector $\big|\psi\big\rangle$ and any $\lambda$;
\item $\mathbf{\hat{E}}(\lambda)\big|\psi\big\rangle \rightarrow 0$ as $\lambda \rightarrow -\infty$ and $\mathbf{\hat{E}}(\lambda)\big|\psi\big\rangle\rightarrow \big|\psi\big\rangle$ as $\lambda\rightarrow \infty$
\end{itemize}
\end{definition}
We will further narrow our focus to Hamiltonians $\mathbf{\hat{H}}$ which are bounded and have spectrum $\sigma(\mathbf{\hat{H}})\subset \mathbb{R}^{+}_{0}$; these are not very restricting conditions but they will allow for things to be a lot clearer henceforth, namely by trivially allowing us to work only with positive Hamiltonians. For a bounded self-adjoint Hamiltonians, focusing on the poisitve ones is not a limitation since a self-adjoint operator $\mathbf{\hat{A}}$ and any spectral shift of said operator $\mathbf{\hat{A}}+\mu\mathbf{\hat{I}}$ generate the same quantum dynamics for all $\mu\in \mathbb{R}$, i.e.  
\begin{equation}
e^{-it(\mathbf{\hat{A}}+\mu\mathbf{\hat{I}})}\boldsymbol{\hat{\rho}}_{0} e^{it(\mathbf{\hat{A}}+\mu\mathbf{\hat{I}})} =
e^{-it\mathbf{\hat{A}}}e^{-it\mu\mathbf{\hat{I}}}\boldsymbol{\hat{\rho}}_{0} e^{it\mathbf{\hat{A}}}e^{it\mu\mathbf{\hat{I}}} =
\end{equation}
\begin{equation}
e^{-it\mu\mathbf{\hat{I}}}e^{it\mu\mathbf{\hat{I}}}e^{-it\mathbf{\hat{A}}}\boldsymbol{\hat{\rho}}_{0} e^{it\mathbf{\hat{A}}} = e^{-it\mathbf{\hat{A}}}\boldsymbol{\hat{\rho}}_{0} e^{it\mathbf{\hat{A}}} 
\end{equation}
Hence, the a bounded self-adjoint operator $\mathbf{\hat{A}}$ with $\sigma(\mathbf{\hat{A}})\subseteq [ -\|\mathbf{\hat{A}}\|,\hspace{2mm} \|\mathbf{\hat{A}}\|]$ and $\mathbf{\hat{A}}+\|\mathbf{\hat{A}}\|\mathbf{\hat{I}}$ with $\sigma(\mathbf{\hat{A}}+\|\mathbf{\hat{A}}\|\mathbf{\hat{I}})\subseteq \big[ 0 ,\hspace{2mm} 2\|\mathbf{\hat{A}}\|\big]$  encapsulate the same physics. Where $\mathbf{\hat{A}}+\|\mathbf{\hat{A}}\|\mathbf{\hat{I}}$ is a positive semi-definite operator. 

Now, using the functional calculus of spectral theory \cite{Simon} we know that
\begin{equation}
e^{-it\mathbf{\hat{H}}} = \int_{\sigma(\mathbf{\hat{H}})}e^{-it\lambda}\mathbf{\hat{E}}(d\lambda)
\end{equation}
 Consequently, for a density operator $\boldsymbol{\hat{\rho}}_{0}$,
\begin{equation}
\bigg\langle\!\!\!\bigg\langle 
e^{-t\mathfrak{\hat{O}}_{\mathbf{\hat{H}}}}\boldsymbol{\hat{\rho}}_{0}\bigg\rangle\!\!\!\bigg\rangle_{T} =   \Bigg\langle \!\!\!\Bigg\langle \bigg(\int_{\sigma(\mathbf{\hat{H}})}e^{-it\lambda}\mathbf{\hat{E}}(d\lambda) \bigg)\boldsymbol{\hat{\rho}}_{0} \bigg(\int_{\sigma(\mathbf{\hat{H}})}e^{it\lambda}\mathbf{\hat{E}}(d\lambda)\bigg)\Bigg\rangle\!\!\!\Bigg\rangle_{T}   = 
\end{equation}
\begin{equation}
\Bigg\langle\!\!\!\Bigg\langle\int_{\sigma(\mathbf{\hat{H}})}\int_{\sigma(\mathbf{\hat{H}})}e^{-it(\mu-\lambda)}\mathbf{\hat{E}}(d\mu) \boldsymbol{\hat{\rho}}_{0} \mathbf{\hat{E}}(d\lambda)\Bigg\rangle\!\!\!\Bigg\rangle_{T} =\int_{\sigma(\mathbf{\hat{H}})}\int_{\sigma(\mathbf{\hat{H}})}\Big\langle\!\!\!\Big\langle e^{-it(\mu-\lambda)}\Big\rangle\!\!\!\Big\rangle_{T}\mathbf{\hat{E}}(d\mu) \boldsymbol{\hat{\rho}}_{0} \mathbf{\hat{E}}(d\lambda)
\end{equation}
Moreover, introducing a resolution limit $\Delta$, we may partition the last integral as follows 
\begin{equation}
\label{eqn:aboutdeltandsig}
\sum_{i}\sum_{j}\int_{\Delta_{i}}\int_{\Delta_{i}}\Big\langle\!\!\!\Big\langle e^{-it(\mu-\lambda)}\Big\rangle\!\!\!\Big\rangle_{T}\mathbf{\hat{E}}(d\mu) \boldsymbol{\hat{\rho}}_{0} \mathbf{\hat{E}}(d\lambda)
\end{equation}
where all of the $\Delta_{i}\subset \sigma(\mathbf{\hat{H}})$ will be assumed to have measure $\Delta$; for this to workout nicely we assume that $\Delta$ divides $|\sigma(\mathbf{\hat{H}})|$. Of course these constraints are simply implemented to make things as simple as possible; distilling the primary focus of this paper,the study equilibration processes for the case of continuous, from less interesting technical challenges. Although the nature of the partition is not the most import thing, a partition must nevertheless take place for us to control decoherence subsectors for generalized pointer bases \cite{zeh}. The latter is so due to the fact that full dephasing is not physically possible in the case of purely continuous variables (i.e. when $\mathbf{\hat{H}}$ has purely continuous spectrum) since this would require $T$ to diverge to infinity, whence the corresponding T-time ergodic averages converge to zero (as was the case in the finite dimensional case \cite{short}); a case that we have already deemed nonphysical. This partition allows us to study coherence subsectors and hence the dephasing with respect to such subsectors \cite{schloss} \cite{zeh}, and in turn to study the dynamics of (\ref{eqn:aboutdeltandsig}). In particular, we will be interested in estimating 
\begin{equation}
\bigg\|\int_{\Delta_{i}}\int_{\Delta_{j}}\Big\langle\!\!\!\Big\langle e^{-it(\mu-\lambda)}\Big\rangle\!\!\!\Big\rangle_{T}\mathbf{\hat{E}}(d\mu) \boldsymbol{\hat{\rho}}_{0} \mathbf{\hat{E}}(d\lambda)\bigg\|_{1}
\end{equation}
\subsection{Preliminary estimates}
\;\;\;Defining $\mathbf{\hat{P}}(\Delta_{i}):=\int_{\Delta_{i}}\mathbf{\hat{E}}(dx)$,  assuming without loss of generality that $\|\mathbf{\hat{A}}\|\leq 1$ and letting $\boldsymbol{\hat{\rho}}_{t}:=e^{-t\mathfrak{\hat{O}}_{\mathbf{\hat{H}}}}\boldsymbol{\hat{\rho}}_{0}$ we now shift our attention to the T-time ergodic average error $\sigma^{2}_{\mathbf{\hat{A}}}(T):= \langle\!\langle|Tr\{\mathbf{\hat{A}}\boldsymbol{\hat{\rho}}_{t}\}- Tr\{\mathbf{\hat{A}}\langle\!\langle\boldsymbol{\hat{\rho}}_{t}\rangle\!\rangle_{T}\}|^{2}\rangle\!\rangle_{T}$ for the case where $\mathbf{\hat{H}}$ has purely continuous spectrum. To that end, we first present the following analysis. 

Define $\mathscr{P}_{\Delta}(\cdot):=\sum_{i}\mathbf{\hat{P}}_{\Delta_{i}}(\cdot)\mathbf{\hat{P}}_{\Delta_{i}}$. Now,
\begin{equation}
\big|Tr\big\{\mathbf{\hat{A}}\boldsymbol{\hat{\rho}}_{t}\big\}- Tr\big\{\mathbf{\hat{A}}\big\langle\!\!\big\langle\boldsymbol{\hat{\rho}}_{t}\big\rangle\!\!\big\rangle_{T}\big\}\big| \leq
\end{equation}
\begin{equation}
\Big|Tr\big\{\mathbf{\hat{A}}\boldsymbol{\hat{\rho}}_{t}\big\}- Tr\Big\{\mathbf{\hat{A}}\sum_{i}\mathbf{\hat{P}}_{\Delta_{i}}\big\langle\!\!\big\langle\boldsymbol{\hat{\rho}}_{t} \big\rangle\!\!\big\rangle_{T}\mathbf{\hat{P}}_{\Delta_{i}}\Big\}-Tr\Big\{\mathbf{\hat{A}}\sum_{i\neq j}\mathbf{\hat{P}}(\Delta_{i})\big\langle\!\!\big\langle\boldsymbol{\hat{\rho}}_{t} \big\rangle\!\!\big\rangle_{T}\mathbf{\hat{P}}(\Delta_{j})\Big\}\Big|\leq
\end{equation}

\begin{equation}
\label{eqn:aptrab0}
\Big|Tr\big\{\mathbf{\hat{A}}\boldsymbol{\hat{\rho}}_{t}\big\}- Tr\Big\{\mathbf{\hat{A}}\sum_{i}\mathbf{\hat{P}}(\Delta_{i})\big\langle\!\!\big\langle\boldsymbol{\hat{\rho}}_{t} \big\rangle\!\!\big\rangle_{T}\mathbf{\hat{P}}(\Delta_{i})\Big\}\Big|+\Big|Tr\Big\{\mathbf{\hat{A}}\sum_{i\neq j}\mathbf{\hat{P}}(\Delta_{i})\big\langle\!\!\big\langle\boldsymbol{\hat{\rho}}_{t} \big\rangle\!\!\big\rangle_{T}\mathbf{\hat{P}}(\Delta_{j})\Big\}\Big|\leq
\end{equation}

\begin{equation}
\label{eqn:aptrb}
\Big|Tr\big\{\mathbf{\hat{A}}\boldsymbol{\hat{\rho}}_{t}\big\}- Tr\Big\{\mathbf{\hat{A}}\sum_{i}\mathbf{\hat{P}}(\Delta_{i})\big\langle\!\!\big\langle\boldsymbol{\hat{\rho}}_{t} \big\rangle\!\!\big\rangle_{T}\mathbf{\hat{P}}(\Delta_{i})\Big\}\Big|+\Big\|\mathbf{\hat{A}}\sum_{i\neq j}\mathbf{\hat{P}}(\Delta_{i})\big\langle\!\!\big\langle\boldsymbol{\hat{\rho}}_{t} \big\rangle\!\!\big\rangle_{T}\mathbf{\hat{P}}(\Delta_{j})\Big\|_{1}\leq
\end{equation}
\begin{equation}
\Big|Tr\big\{\mathbf{\hat{A}}\boldsymbol{\hat{\rho}}_{t}\big\}- Tr\Big\{\mathbf{\hat{A}}\sum_{i}\mathbf{\hat{P}}(\Delta_{i})\big\langle\!\!\big\langle\boldsymbol{\hat{\rho}}_{t} \big\rangle\!\!\big\rangle_{T}\mathbf{\hat{P}}(\Delta_{i})\Big\}\Big|+\sum_{i\neq j}\Big\|\mathbf{\hat{P}}(\Delta_{i})\big\langle\!\!\big\langle\boldsymbol{\hat{\rho}}_{t} \big\rangle\!\!\big\rangle_{T}\mathbf{\hat{P}}(\Delta_{j})\Big\|_{1}\leq
\end{equation}
\begin{equation}
\Big| Tr\Big\{\mathbf{\hat{A}}\sum_{i\neq j}\mathbf{\hat{P}}(\Delta_{i})\boldsymbol{\hat{\rho}}_{t}\mathbf{\hat{P}}(\Delta_{i})\Big\}\Big|+\sum_{i\neq j}\Big\|\mathbf{\hat{P}}(\Delta_{i})\big\langle\!\!\big\langle\boldsymbol{\hat{\rho}}_{t} \big\rangle\!\!\big\rangle_{T}\mathbf{\hat{P}}(\Delta_{j})\Big\|_{1}+
\end{equation}
\begin{equation}
\Big|Tr\Big\{\mathbf{\hat{A}}\sum_{i}\mathbf{\hat{P}}(\Delta_{i})\boldsymbol{\hat{\rho}}_{t} \mathbf{\hat{P}}(\Delta_{i})\Big\}- Tr\Big\{\mathbf{\hat{A}}\sum_{i}\mathbf{\hat{P}}(\Delta_{i})\big\langle\!\!\big\langle\boldsymbol{\hat{\rho}}_{t} \big\rangle\!\!\big\rangle_{T}\mathbf{\hat{P}}(\Delta_{i})\Big\}\Big|
\end{equation}
Here, we have applied the inequalities $|Tr\{\mathbf{\hat{A}}\}|\leq \|\mathbf{\hat{A}}\|_{1}$ \cite{lax},$\|\mathbf{\hat{A}}\mathbf{\hat{B}}\|_{1}\leq\|\mathbf{\hat{A}}\|\|\mathbf{\hat{B}}\|_{1}$ and triangle inequality in going from equation (\ref{eqn:aptrab0}) to equation (\ref{eqn:aptrb}). Hence, using the fact that $\big(\sum_{i=1}^{n}x_{i}\big)^{2}\leq n\sum_{i=1}^{n}x_{i}^{2}$ 
\begin{equation}
\frac{1}{3}\sigma^{2}_{\mathbf{\hat{A}}}(T):=\frac{1}{3}\bigg\langle\!\!\!\bigg\langle\big|Tr\big\{\mathbf{\hat{A}}\boldsymbol{\hat{\rho}}_{t}\big\}- Tr\big\{\mathbf{\hat{A}}\big\langle\!\!\big\langle\boldsymbol{\hat{\rho}}_{t} \big\rangle\!\!\big\rangle_{T}\big\}\big|^{2}\bigg\rangle\!\!\!\bigg\rangle_{T}\leq 
\end{equation}
\begin{equation}
\Bigg\langle\!\!\!\Bigg\langle\Big| Tr\Big\{\mathbf{\hat{A}}\sum_{i\neq j}\mathbf{\hat{P}}(\Delta_{i})\boldsymbol{\hat{\rho}}_{t}\mathbf{\hat{P}}(\Delta_{j})\Big\}\Big|^{2}\Bigg\rangle\!\!\!\Bigg\rangle_{T}+\bigg(\sum_{i\neq j}\Big\|\mathbf{\hat{P}}(\Delta_{i})\big\langle\!\!\big\langle\boldsymbol{\hat{\rho}}_{t} \big\rangle\!\!\big\rangle_{T}\mathbf{\hat{P}}(\Delta_{j})\Big\|_{1}\bigg)^{2}+
\end{equation}
\begin{equation}
\Bigg\langle\!\!\!\Bigg\langle\Big|Tr\Big\{\mathbf{\hat{A}}\sum_{i}\mathbf{\hat{P}}(\Delta_{i})\boldsymbol{\hat{\rho}}_{t} \mathbf{\hat{P}}(\Delta_{i})\Big\}- Tr\Big\{\mathbf{\hat{A}}\sum_{i}\mathbf{\hat{P}}(\Delta_{i})\big\langle\!\!\big\langle\boldsymbol{\hat{\rho}}_{t} \big\rangle\!\!\big\rangle_{T} \mathbf{\hat{P}}(\Delta_{i})\Big\}\Big|^{2}\Bigg\rangle\!\!\!\Bigg\rangle_{T}
\end{equation}
Now, unlike the proof from \cite{short} leading to Theorem \ref{eqn:themainresultriemann}, here we have three terms that must be analyzed instead of only one. For the sake of clarity, let us give these terms names  
\begin{equation}
\mathscr{R}_{\mathbf{\hat{A}}}(T,\Delta):=\Bigg\langle\!\!\!\Bigg\langle\Big| Tr\Big\{\mathbf{\hat{A}}\sum_{i\neq j}\mathbf{\hat{P}}(\Delta_{i})\boldsymbol{\hat{\rho}}_{t}\mathbf{\hat{P}}(\Delta_{j})\Big\}\Big|^{2}\Bigg\rangle\!\!\!\Bigg\rangle_{T}
\end{equation}
\begin{equation}
\mathscr{D}_{\mathbf{\hat{A}}}(T,\Delta):=
\Bigg\langle\!\!\!\Bigg\langle\Big|Tr\Big\{\mathbf{\hat{A}}\mathscr{P}_{\Delta}\big(\boldsymbol{\hat{\rho}}_{t}\big)\Big\}- Tr\Big\{\mathbf{\hat{A}}\mathscr{P}_{\Delta}\big(\big\langle\!\!\big\langle\boldsymbol{\hat{\rho}}_{t} \big\rangle\!\!\big\rangle_{T}\big)\Big\}\Big|^{2}\Bigg\rangle\!\!\!\Bigg\rangle_{T}   
\end{equation}
\begin{equation}
\label{eqn:kupchbounding}
\mathscr{K}(T,\Delta):= \bigg(\sum_{i\neq j}\Big\|\mathbf{\hat{P}}(\Delta_{i})\big\langle\!\!\big\langle\boldsymbol{\hat{\rho}}_{t} \big\rangle\!\!\big\rangle_{T}\mathbf{\hat{P}}(\Delta_{j})\Big\|_{1}\bigg)^{2} \end{equation}
\subsection{A bound for estimating $\mathscr{K}(T, \Delta)$ and $\mathscr{D}_{\mathbf{\hat{A}}}(T,\Delta)$}
\;\;\; For the case where $\mathbf{\hat{H}}$ has purely point spectrum, e.g.\cite{short}, where letting $T\rightarrow \infty$ is physically viable, there are no terms analogous to $\mathscr{K}(T,\Delta)$ and $\mathscr{D}_{\mathbf{\hat{A}}}(T,\Delta)$. These terms encapsulate the coherences present after the finite equilibration time $T$ has elapsed and the relative invariance of the dynamics pertaining to the subspace associated with the projectors $\mathbf{\hat{P}}(\Delta_{i})$ respectively; i.e. $\mathbf{\hat{P}}(\Delta_{i})\boldsymbol{\hat{\rho}}_{t} \mathbf{\hat{P}}(\Delta_{i})$. To estimate the term $\mathscr{K}(T,\Delta)$, we may modify a result in \cite{acevedo3}, which studies a similar object albeit in a more constrained setting. We present this result as a lemma below.
\begin{definition2}[\textbf{Coherence subsector estimates }]
\label{eqn:theoremkupsch}
Let us fix $t>0$ and let $\boldsymbol{\hat{\rho}}_{0}$ be a density operator. Furthermore, let $\Gamma(t,x,y)$ be a smooth function over a domain $\Omega$ which is symmetric with respect to $x,$ and $y$, and let $\mathbf{\hat{E}}(x)$ be a spectral family of projection operators and define $\mathbf{\hat{P}}(\Delta_{i}):=\int_{\Delta_{i}}\mathbf{\hat{E}}(dx)$, $\Delta_{i}\subset \Omega$. Finally, define
\begin{equation}
\boldsymbol{\hat{\rho}}_{t} := \int\int  \Gamma(t,x,y)\mathbf{\hat{E}}(dx)\boldsymbol{\hat{\rho}}_{0}\mathbf{\hat{E}}(dy)
\end{equation}
Then, for $i\neq j$,
\begin{equation}
\label{eqn:kupschkupsch}
\big\|\mathbf{\hat{P}}(\Delta_{i})\boldsymbol{\hat{\rho}}_{t}\mathbf{\hat{P}}(\Delta_{j})\big\|_{1}\leq \sup_{(x,y)\in \Delta_{i}\times \Delta_{j}}\bigg(2|\Gamma(t,x,y)|+|\Delta_{j}||\partial_{y}\Gamma(t,x,y)|\bigg)
\end{equation}
when $\sup\{\boldsymbol{\hat{\rho}}_{0}\}\cap \sup\{\mathbf{\hat{P}}(\Delta_{i})\} \neq  0$ and $\sup\{\boldsymbol{\hat{\rho}}_{0}\}\cap \sup\{\mathbf{\hat{P}}(\Delta_{j})\} \neq 0$;
otherwise, 
\begin{equation}
\big\|\mathbf{\hat{P}}(\Delta_{i})\boldsymbol{\hat{\rho}}_{t}\mathbf{\hat{P}}(\Delta_{j})\big\|_{1} = 0
\end{equation}
\end{definition2}
\begin{proof}
 The proof may be found in appendix \ref{eqn:appa}.   
\end{proof}
\subsection{Estimates for  $\mathscr{D}_{\mathbf{\hat{A}}}(T,\Delta)$ and $\mathscr{R}_{\mathbf{\hat{A}}}(T,\Delta)$}
\;\;\; We now analyze the term $\mathscr{D}_{\mathbf{\hat{A}}}(T,\Delta)$. With some effort, the following lemma may be proven.
\begin{definition2}[\textbf{Estimates for} $\mathscr{D}_{\mathbf{\hat{A}}}(T,\Delta)$ ]
\label{eqn:lemma2}
Given the assumptions leading to the definition of $\mathscr{D}_{\mathbf{\hat{A}}}(T,\Delta)$, the following result holds
\small
\begin{equation}
\label{eqn:coke}
\mathscr{D}_{\mathbf{\hat{A}}}(T,\Delta)\leq 4\max_{i,j}\sup_{\tau\in[-T,T]}\bigg(1-\beta_{ij}^{-2}\Big|\int_{\Delta_{i}}e^{-i\tau x}\mu_{\psi_{j}}(dx)\Big|^{2}\bigg)
\end{equation}
where the measure $\mu_{\psi_{j}}(dx):=\big\langle \psi_{j}\big|\mathbf{\hat{E}}(dx)\big|\psi_{j}\big\rangle$ and the $\big|\psi_{j}\big\rangle$ are implicitly defined via the state at time zero $\boldsymbol{\hat{\rho}}_{0}:=\sum_{j}\alpha_{j}\big|\psi_{j}\big\rangle\big\langle\psi_{j}\big|$. Furthermore $\beta_{ij}:= \big\langle \psi_{j}\big|\mathbf{\hat{P}}(\Delta_{i})\big|\psi_{j}\big\rangle$.
\end{definition2}
\begin{proof}
To see the proof of this the reader is referred to appendix \ref{eqn:appb}.
\end{proof}
To avoid clutter, we will use the following definition in what follows.

\begin{equation}
\mathscr{F}(T,\Delta):=4\max_{i,j}\sup_{\tau\in[-T,T]}\bigg(1-\beta_{ij}^{-2}\Big|\int_{\Delta_{i}}e^{-i\tau x}\mu_{\psi_{j}}(dx)\Big|^{2}\bigg)
\end{equation}

Finally, we analyze the term $\mathscr{R}_{\mathbf{\hat{A}}}(T,\Delta)$. However, just as it was done in \cite{short}, here we shall assume that $\boldsymbol{\hat{\rho}}_{0}$ is a pure state $|\psi(0)\rangle\langle\psi(0)|$; we do this because obtaining estimates for the general case of a mixture remains an elusive task for the term $\mathscr{R}_{\mathbf{\hat{A}}}(T,\Delta)$. Let us now consider the following sequence of equalities and inequalities. 
\begin{equation}
\Bigg\langle\!\!\!\Bigg\langle\Big| Tr\Big\{\mathbf{\hat{A}}\sum_{i\neq j}\mathbf{\hat{P}}(\Delta_{i})\boldsymbol{\hat{\rho}}_{t}\mathbf{\hat{P}}(\Delta_{j})\Big\}\Big|^{2}\Bigg\rangle\!\!\!\Bigg\rangle_{T} = 
\end{equation}
\begin{equation}
\label{eqn:pen4}
\Bigg\langle\!\!\!\Bigg\langle\sum_{i\neq j}\sum_{l\neq k}Tr\Big\{\mathbf{\hat{A}}\mathbf{\hat{P}}(\Delta_{i})\boldsymbol{\hat{\rho}}_{t}\mathbf{\hat{P}}(\Delta_{j})\Big\} Tr\Big\{\mathbf{\hat{A}}^{\dagger}\mathbf{\hat{P}}(\Delta_{l})\boldsymbol{\hat{\rho}}_{t}\mathbf{\hat{P}}(\Delta_{k})\Big\}\Bigg\rangle\!\!\!\Bigg\rangle_{T} =   
\end{equation}
\begin{equation}
\sum_{i\neq j}\sum_{l\neq k}\bigg\langle\!\!\!\bigg\langle Tr\Big\{\mathbf{\hat{A}}\mathbf{\hat{P}}(\Delta_{i})\boldsymbol{\hat{\rho}}_{t}\mathbf{\hat{P}}(\Delta_{j})\otimes \mathbf{\hat{A}}^{\dagger}\mathbf{\hat{P}}(\Delta_{l})\boldsymbol{\hat{\rho}}_{t}\mathbf{\hat{P}}(\Delta_{k})\Big\} \bigg\rangle\!\!\!\bigg\rangle_{T} =  
\end{equation}
\begin{equation}
\sum_{i\neq j}\sum_{l\neq k}\bigg\langle\!\!\!\bigg\langle Tr\Big\{\Big(\mathbf{\hat{A}}\mathbf{\hat{P}}(\Delta_{i})\otimes\mathbf{\hat{A}}^{\dagger}\mathbf{\hat{P}}(\Delta_{l})\Big)\Big(\boldsymbol{\hat{\rho}}_{t}\otimes \boldsymbol{\hat{\rho}}_{t}\Big)\Big(\mathbf{\hat{P}}(\Delta_{j})\otimes\mathbf{\hat{P}}(\Delta_{k})\Big)\Big\} \bigg\rangle\!\!\!\bigg\rangle_{T} 
\end{equation}
\begin{equation}
\sum_{i\neq j}\sum_{l\neq k} Tr\Big\{\Big(\mathbf{\hat{A}}\otimes\mathbf{\hat{A}}^{\dagger}\Big)\big(\mathbf{\hat{P}}(\Delta_{i})\otimes\mathbf{\hat{P}}(\Delta_{l})\Big)\mathscr{E}_{T}\Big(\boldsymbol{\hat{\rho}}_{0}\otimes \boldsymbol{\hat{\rho}}_{0}\Big)\Big(\mathbf{\hat{P}}(\Delta_{j})\otimes\mathbf{\hat{P}}(\Delta_{k})\Big)\Big\} \leq 
\end{equation}
\begin{equation}
\sum_{i\neq j}\sum_{l\neq k} \Big\|\Big(\mathbf{\hat{P}}(\Delta_{i})\otimes\mathbf{\hat{P}}(\Delta_{l})\Big)\mathscr{E}_{T}\Big(\boldsymbol{\hat{\rho}}_{0}\otimes \boldsymbol{\hat{\rho}}_{0}\Big)\Big(\mathbf{\hat{P}}(\Delta_{j})\otimes\mathbf{\hat{P}}(\Delta_{k})\Big)\Big\|_{1} =
\end{equation}
\begin{equation}
\label{eqn:kup2}
\sum_{i\neq j}\sum_{l\neq k} \Big\|\mathbf{\hat{G}}(\Delta_{i},\Delta_{l})\mathscr{E}_{T}\big(\boldsymbol{\hat{\sigma}}_{0}\big)\mathbf{\hat{G}}(\Delta_{j}, \Delta_{k})\Big\|_{1}
\end{equation}
Here we have made the following definitions.
\begin{itemize}
\item $\mathbf{\hat{G}}(\Delta_{n},\Delta_{m}):= \mathbf{\hat{P}}(\Delta_{n})\otimes\mathbf{\hat{P}}(\Delta_{m})$
\item$\mathscr{E}_{T}\big(\cdot\big):=\frac{1}{T}\int_{0}^{T}e^{-it\mathbf{\hat{H}}}\otimes e^{-it\mathbf{\hat{H}}}\big(\cdot\big)e^{it\mathbf{\hat{H}}}\otimes e^{it\mathbf{\hat{H}}}dt = $\\

$\int\int\int\int\big\langle\!\!\big\langle e^{-it(x-y+w-v)}\big\rangle\!\!\big\rangle_{T}\mathbf{\hat{E}}(dx)\otimes \mathbf{\hat{E}}(dw)\big(\cdot\big)\mathbf{\hat{E}}(dy)\otimes \mathbf{\hat{E}}(dv) $

\item $\boldsymbol{\hat{\sigma}}_{0}:= \boldsymbol{\hat{\rho}}_{0}\otimes \boldsymbol{\hat{\rho}}_{0}$ 
\end{itemize}
The elements of the sum $(\ref{eqn:kup2})$ are indeed of the type already encountered when discussing the term $\mathscr{K}(T,\Delta)$, namely, in the scheme subsequently presented to estimate such terms in Lemma \ref{eqn:theoremkupsch}. For the terms in the sum (\ref{eqn:kup2}) where $i\neq l$ or $j\neq k$, we may estimate the $T$ and $\Delta$ dependent decay by applying Lemma \ref{eqn:theoremkupsch}. This may be seen by noting that 
\begin{equation}
\mathbf{\hat{G}}(\Delta_{i},\Delta_{l})\mathscr{E}_{T}\big(\boldsymbol{\hat{\sigma}}_{0}\big)\mathbf{\hat{G}}(\Delta_{j}, \Delta_{k}) = 
\int_{\Delta_{i}}\int_{\Delta_{j}}\int_{\Delta_{l}}\int_{\Delta_{k}}\big\langle\!\!\big\langle e^{-it(x-y+w-v)}\big\rangle\!\!\big\rangle_{T}\mathbf{\hat{E}}(dx)\otimes \mathbf{\hat{E}}(dw)\big(\boldsymbol{\hat{\sigma}}_{0}\big)\mathbf{\hat{E}}(dy)\otimes \mathbf{\hat{E}}(dv) 
\end{equation}
where the order of integration is $dxdydwdv$. The \emph{decoherence kernel} \cite{schloss} $\langle\!\langle e^{-it(x-y+w-v)}\rangle\!\rangle_{T}$ can be small only if the term $x-y+w-v$ is large. However, for the case where $i=l$ and $j = k$ the term $\langle\!\langle e^{-it(x-y+w-v)}\rangle\!\rangle_{T}$ is expected to be approximately 1, whence other means of estimation are needed for such terms. To address the case where $i = k$ and $j = l$ we start our estimation from equation (\ref{eqn:pen4}). 
\begin{equation}
\Bigg\langle\!\!\!\Bigg\langle\sum_{i\neq j}Tr\Big\{\mathbf{\hat{A}}\mathbf{\hat{P}}(\Delta_{i})\boldsymbol{\hat{\rho}}(t)\mathbf{\hat{P}}(\Delta_{j})\Big\} Tr\Big\{\mathbf{\hat{A}}^{\dagger}\mathbf{\hat{P}}(\Delta_{j})\boldsymbol{\hat{\rho}}(t)\mathbf{\hat{P}}(\Delta_{i})\Big\}\Bigg\rangle\!\!\!\Bigg\rangle_{T} =
\end{equation}
\begin{equation}
\Bigg\langle\!\!\!\Bigg\langle\sum_{i\neq j}\big\langle \psi(t)\big|\mathbf{\hat{P}}(\Delta_{j})\mathbf{\hat{A}}\mathbf{\hat{P}}(\Delta_{i})\big|\psi(t)\big\rangle \big\langle \psi(t)\big|\mathbf{\hat{P}}(\Delta_{i})\mathbf{\hat{A}}^{\dagger}\mathbf{\hat{P}}(\Delta_{j})\big|\psi(t)\big\rangle\Bigg\rangle\!\!\!\Bigg\rangle_{T} = 
\end{equation}
\begin{equation}
\Bigg\langle\!\!\!\Bigg\langle\sum_{i\neq j}Tr\Big\{\mathbf{\hat{A}}\mathbf{\hat{P}}(\Delta_{i})\big|\psi(t)\big\rangle \big\langle \psi(t)\big|\mathbf{\hat{P}}(\Delta_{i})\mathbf{\hat{A}}^{\dagger}\mathbf{\hat{P}}(\Delta_{j})\big|\psi(t)\big\rangle\big\langle \psi(t)\big|\mathbf{\hat{P}}(\Delta_{j})\Big\}\Bigg\rangle\!\!\!\Bigg\rangle_{T} \leq
\end{equation}
\begin{equation}
\Bigg\langle\!\!\!\Bigg\langle\sum_{i}\sum_{j}Tr\Big\{\mathbf{\hat{A}}\mathbf{\hat{P}}(\Delta_{i})\big|\psi(t)\big\rangle \big\langle \psi(t)\big|\mathbf{\hat{P}}(\Delta_{i})\mathbf{\hat{A}}^{\dagger}\mathbf{\hat{P}}(\Delta_{j})\big|\psi(t)\big\rangle\big\langle \psi(t)\big|\mathbf{\hat{P}}(\Delta_{j})\Big\}\Bigg\rangle\!\!\!\Bigg\rangle_{T} =
\end{equation}
\begin{equation}
\Bigg\langle\!\!\!\Bigg\langle Tr\Big\{\mathbf{\hat{A}}\mathscr{P}_{\Delta}\Big(\big|\psi(t)\big\rangle \big\langle \psi(t)\big|\Big)\mathbf{\hat{A}}^{\dagger}\mathscr{P}_{\Delta}\Big(\big|\psi(t)\big\rangle\big\langle \psi(t)\big|\Big)\Big\}\Bigg\rangle\!\!\!\Bigg\rangle_{T} \leq
\end{equation}
\begin{equation}
\Bigg\langle\!\!\!\Bigg\langle\sqrt{ Tr\Big\{\mathbf{\hat{A}}^{\dagger}\mathbf{\hat{A}}\mathscr{P}_{\Delta}\Big(\big|\psi(t)\big\rangle \big\langle \psi(t)\big|\Big)^{2}\Big\}Tr\Big\{\mathbf{\hat{A}}\mathbf{\hat{A}}^{\dagger}\mathscr{P}_{\Delta}\Big(\big|\psi(t)\big\rangle \big\langle \psi(t)\big|\Big)^{2}\Big\}}\Bigg\rangle\!\!\!\Bigg\rangle_{T} \leq 
\end{equation}
\begin{equation}
\|\mathbf{\hat{A}}\|^{2}\Bigg\langle\!\!\!\Bigg\langle\sqrt{ Tr\Big\{\mathscr{P}_{\Delta}\Big(\big|\psi(t)\big\rangle \big\langle \psi(t)\big|\Big)^{2}\Big\}Tr\Big\{\mathscr{P}_{\Delta}\Big(\big|\psi(t)\big\rangle \big\langle \psi(t)\big|\Big)^{2}\Big\}}\Bigg\rangle\!\!\!\Bigg\rangle_{T} =
\end{equation}
\begin{equation}
\Bigg\langle\!\!\!\Bigg\langle  Tr\Big\{\mathscr{P}_{\Delta}\Big(\big|\psi(t)\big\rangle \big\langle \psi(t)\big|\Big)^{2}\Big\}\Bigg\rangle\!\!\!\Bigg\rangle _{T} =
\Bigg\langle\!\!\!\Bigg\langle Tr\Big\{e^{-it\mathbf{\hat{H}}}\mathscr{P}_{\Delta}\Big(\big|\psi(0)\big\rangle \big\langle \psi(0)\big|\Big)^{2}e^{it\mathbf{\hat{H}}}\Big\}\Bigg\rangle\!\!\!\Bigg\rangle_{T} =
\end{equation}
\begin{equation}
Tr\Big\{\mathscr{P}_{\Delta}\Big(\big|\psi(0)\big\rangle \big\langle \psi(0)\big|\Big)^{2}\Big\}=
\end{equation}
\begin{equation}
\sum_{i}Tr\Big\{\mathbf{\hat{P}}(\Delta_{i})\big|\psi(0)\big\rangle \big\langle \psi(0)\big|\mathbf{\hat{P}}(\Delta_{i})\big|\psi(0)\big\rangle \big\langle \psi(0)\big|\mathbf{\hat{P}}(\Delta_{i})\Big\} =  
\end{equation}
\begin{equation}
\sum_{i}\big\langle \psi(0)\big|\mathbf{\hat{P}}(\Delta_{i})\big|\psi(0)\big\rangle Tr\Big\{\mathbf{\hat{P}}(\Delta_{i})\big|\psi(0)\big\rangle \big\langle \psi(0)\big|\mathbf{\hat{P}}(\Delta_{i})\Big\} =
\end{equation}
\begin{equation}
\sum_{i}\big\langle \psi(0)\big|\mathbf{\hat{P}}(\Delta_{i})\big|\psi(0)\big\rangle^{2} \end{equation}
Note the similarity between this estimate and the estimate provided by Short in Theorem 1 of \cite{short} for the finite-dimensional case; this is indeed an inverted effective dimension of sorts. We present these latter results as a lemma in the box below.  
\begin{definition2}[\textbf{Estimates for} $\mathscr{R}_{\mathbf{\hat{A}}}(T,\Delta)$ ]
\label{eqn:lem3}
Given the assumptions leading to the definition of $\mathscr{R}_{\mathbf{\hat{A}}}(T,\Delta)$, and assuming that $\boldsymbol{\hat{\rho}}_{0} = |\psi(0)\rangle\langle\psi(0)|$, the following holds.
\small
\begin{equation}
\mathscr{R}_{\mathbf{\hat{A}}}(T,\Delta) \leq \sum_{\overset{i\neq j;l\neq k}{i\neq l\hspace{1mm} or \hspace{1mm} j \neq k}}\Big\|\mathbf{\hat{G}}(\Delta_{i},\Delta_{l})\mathscr{E}_{T}\big(\boldsymbol{\hat{\sigma}}_{0}\big)\mathbf{\hat{G}}(\Delta_{j}, \Delta_{k})\Big\|_{1}+\sum_{i}\big\langle \psi(0)\big|\mathbf{\hat{P}}(\Delta_{i})\big|\psi(0)\big\rangle^{2} 
\end{equation}
\normalsize
where 
\begin{equation}
\mathbf{\hat{G}}(\Delta_{i},\Delta_{l}):= \mathbf{\hat{P}}(\Delta_{i})\otimes\mathbf{\hat{P}}(\Delta_{l})
\end{equation}
\begin{equation}
\mathscr{E}_{T}(\cdot):=\frac{1}{T}\int_{0}^{T}e^{-it\mathbf{\hat{H}}}\otimes e^{-it\mathbf{\hat{H}}}(\cdot)e^{it\mathbf{\hat{H}}}\otimes e^{it\mathbf{\hat{H}}}dt
\end{equation}
\begin{equation}
\boldsymbol{\hat{\sigma}}_{0}:= \boldsymbol{\hat{\rho}}_{0}\otimes \boldsymbol{\hat{\rho}}_{0}
\end{equation}
\end{definition2}
\begin{proof}
See the discussion preceding Lemma \ref{eqn:lem3}
\end{proof}
\subsection{The main result}
\;\;\;We conclude this section with a theorem for the case of continuous variables analogous to Theorem \ref{eqn:themainresultriemann} from \cite{short}; i.e. for the case of infinite dimensional Hilbert spaces and a Hamiltonian $\mathbf{\hat{H}}$ with purely continuous spectrum.
\begin{definition4}
\label{eqn:themain}
Consider an infinite dimensional Hilbert space $\mathscr{H}$ and a  Hamiltonian $\mathbf{\hat{H}} \in L(\mathscr{H})$ with purely continuous spectrum and with spectral decomposition $\mathbf{\hat{H}} = \int_{\sigma(\mathbf{\hat{H}})}\lambda \mathbf{\hat{E}}(d\lambda)$. Furthermore, let $\boldsymbol{\hat{\rho}}_{t}:=e^{-t\mathfrak{\hat{O}}_{\mathbf{\hat{H}}}}|\psi(0)\rangle\langle\psi(0)|=e^{-it\mathbf{\hat{H}}}|\psi(0)\rangle\langle\psi(0)|e^{it\mathbf{\hat{H}}}$ and denote the $T$-time average by $\langle\!\langle \boldsymbol{\hat{\rho}}_{t}\rangle\!\rangle_{T}$. Finally, let $\Delta_{i}\subset \sigma(\mathbf{\hat{H}})$ with $|\Delta_{i}|=\Delta$ for all $i$, where it is assumed that $\Delta$ divides $|\sigma(\mathbf{\hat{H}})|$ (see the discussion following \ref{eqn:aboutdeltandsig}). Then, 
\begin{equation}
\frac{1}{3}\sigma_{\mathbf{\hat{A}}}^{2}(T):=\frac{1}{3}\Big\langle\!\!\!\Big\langle\big|Tr\big\{\mathbf{\hat{A}}\boldsymbol{\hat{\rho}}_{t}\big\}- Tr\big\{\mathbf{\hat{A}}\big\langle\!\!\big\langle\boldsymbol{\hat{\rho}}_{t}\big\rangle\!\!\big\rangle_{T}\big\}\big|^{2} \Big\rangle\!\!\!\Big\rangle_{T}\leq 
\end{equation}
\begin{equation}
\label{eqn:boig1}
\Big(\sum_{\mathcal{J}_{ij}}\sup_{\omega_{ij}}F(T,\Delta,x,y)\Big)^{2} + \mathscr{F}(T,\Delta) +
\sum_{i}\beta_{i}^{2}
\end{equation}
\begin{equation}
\label{eqn:boig2}
\sum_{\mathcal{I}_{ijlk}}\sup_{\Omega_{ijlk}}F(T,\Delta,x-v, y-w) 
\end{equation}
where
\begin{equation}
F(T,\Delta,x,y):=2|sinc((x-y)T/2)|+\Delta|\partial_{y}(e^{-iT(x-y)/2}sinc((x-y)T/2))|
\end{equation}
\begin{equation}
\mu_{\psi}(dx):=\langle \psi(0)|\mathbf{\hat{E}}(dx)|\psi(0)\rangle,\;\;\;
\beta_{i}:= \langle \psi(0)|\mathbf{\hat{P}}(\Delta_{i})|\psi(0)\rangle,\;\;\;
\mathbf{\hat{P}}(\Delta_{i}):=\int_{\Delta_{i}}\mathbf{\hat{E}}(d\lambda)
\end{equation}
\begin{equation}
\mathcal{I}_{ijkl}:=\{(i,j,l,k)|i\neq j;l\neq k;i\neq l\hspace{1mm} or \hspace{1mm} j \neq k\}
\end{equation}
\begin{equation}
\mathcal{I}_{ij}:=\{(i,j)|i\neq j\},\;\;
\omega_{ij}:=\{(x,y)|x\in\Delta_{i};y\in\Delta_{j}\}
\end{equation}
\begin{equation}
\Omega_{ijkl}:=\{(x,y,w,v)|x\in\Delta_{i};y\in\Delta_{j};w\in\Delta_{l};v\in\Delta_{k}\}
\end{equation}
\end{definition4}
\begin{proof}
This result follows almost immediately from Lemmas \ref{eqn:theoremkupsch}, \ref{eqn:lemma2} and \ref{eqn:lem3}. Nevertheless, a few things will need to be elucidated. Firstly, the second term from the left in (\ref{eqn:boig1}) comes about from Lemma \ref{eqn:lemma2} and the fact that $\boldsymbol{\hat{\rho}}_{0}:=|\psi(0)\rangle\langle \psi(0)|$. The first term from the left in (\ref{eqn:boig1}) comes from applying Lemma \ref{eqn:theoremkupsch} to the bound (\ref{eqn:kupchbounding}) and noting that the $\Gamma(t,x,y)$ in Lemma \ref{eqn:theoremkupsch} is equal to $\langle\!\langle e^{-it(x-y)}\rangle\!\rangle_{T} = e^{-iT(x-y)/2}sinc((x-y)T/2)$ in this case. Similarly, one may obtain the term (\ref{eqn:boig2}) by first applying Lemma \ref{eqn:lem3} and then Lemma \ref{eqn:theoremkupsch}. Finally, the third term from the left of (\ref{eqn:boig1}) follows immediately from \ref{eqn:lem3}. 
\end{proof}
In general, it is difficult to derive the spectral family of projectors characterizing the spectral decomposition of a Hamiltonian  $\mathbf{\hat{H}}$. However, in cases where $\mathbf{\hat{H}}$ is equal to the composition of a continuous function and a quadrature operator \cite{knight}, an example of which is the momentum or position operator. Let $\mathbf{\hat{P}}$ be the canonical momentum operator and let $\mathbf{\hat{H}} = f(\mathbf{\hat{P}})$ ($f(x): \mathbb{R}\rightarrow \mathbb{R}$, a smooth function), we may use the functional calculus to express $\mathbf{\hat{H}}$ as follows.
\begin{equation}
f(\mathbf{\hat{P}})= f\Big(\int_{\sigma(\mathbf{\hat{P}})}p\mathbf{\hat{E}}(dp)\Big)=\int_{\sigma(\mathbf{\hat{P}})}f(p)\mathbf{\hat{E}}(dp)
\end{equation}
The mathematical properties of the spectral measure $\mathbf{\hat{E}}(dp)$ characterizing the spectral decomposition of the momentum operator are known from quantum mechanics; the mathematical properties for any other quadrature operator are identical. The physics convention is two write $\mathbf{\hat{E}}(dp) = |p\rangle\langle p| dp$ where $|p\rangle$ is the so-called generalized eigenvectors of the momentum operator corresponding to the element $p\in \sigma(\mathbf{\hat{P}})$ of the spectrum \cite{sakurai}\cite{pascual}; these are indeed just delta distributions $\langle p| p^{'}\rangle = \delta(p-p^{'})$. Using this notation, $f(\mathbf{\hat{P}}) = \int_{\sigma(\mathbf{\hat{P}})}f(p)|p\rangle\langle p| dp $, and therefore $e^{-it f(\mathbf{\hat{P}})} = \int_{\sigma(\mathbf{\hat{P}})}e^{-itf(p)}|p\rangle\langle p| dp $. We may arrive at a more specialized version of Theorem \ref{eqn:themain} with the latter. Namely, the corollary that follows.
\begin{Co}
\label{eqn:themain2}
Consider an infinite dimensional Hilbert space $\mathscr{H}$ and a Hamiltonian in $ L(\mathscr{H})$ of the form $f(\mathbf{\hat{P}}) = \int_{\sup f(x)}f(p)|p\rangle\langle p| dp$. Here, $\mathbf{\hat{P}}$ is a quadrature operator. Furthermore, let $\boldsymbol{\hat{\rho}}_{t}:=e^{-itf(\mathbf{\hat{P}})}|\psi\rangle\langle\psi|e^{itf(\mathbf{\hat{P}})}$ and denote the $T$-time average by $\langle\!\langle \boldsymbol{\hat{\rho}}_{t}\rangle\!\rangle_{T}$. Finally, let $\Delta_{i}\subset \sup f(x)$ with $|\Delta_{i}|=\Delta$ for all $i$, where it is assumed that $\Delta$ divides $|\sup f(x)|$.
 
\begin{equation}
\frac{1}{3}\sigma_{\mathbf{\hat{A}}}^{2}(T):=\frac{1}{3}\Big\langle\!\!\!\Big\langle\big|Tr\big\{\mathbf{\hat{A}}\boldsymbol{\hat{\rho}}_{t}\big\}- Tr\big\{\mathbf{\hat{A}}\big\langle\!\!\big\langle\boldsymbol{\hat{\rho}}_{t}\big\rangle\!\!\big\rangle_{T}\big\}\big|^{2} \Big\rangle\!\!\!\Big\rangle_{T}\leq 
\end{equation}
\begin{equation}
\Big(\sum_{\mathcal{J}_{ij}}\sup_{\omega_{ij}}F(T,\Delta,f(x),f(y))\Big)^{2} + \mathscr{F}(T,\Delta) +
\sum_{i}\beta_{i}^{2}
\end{equation}
\begin{equation}
\sum_{\mathcal{I}_{ijlk}}\sup_{\Omega_{ijlk}}F(T,\Delta,f(x)-f(v), f(y)-f(w)) 
\end{equation}
where
\begin{equation}
F(T,\Delta,x,y):=2|sinc((x-y)T/2)|+\Delta|\partial_{y}(e^{-iT(x-y)/2}sinc((x-y)T/2))|
\end{equation}
\begin{equation}
\mu_{\psi}(dx):=\langle \psi(0)|\mathbf{\hat{E}}(dx)|\psi(0)\rangle,\;\;\;
\beta_{i}:= \int_{\Delta_{i}}|\psi_{i}(p)|^{2}dp,\;\;\;
\mathbf{\hat{P}}(\Delta_{i}):=\int_{\Delta_{i}}|p\rangle\langle p|dp
\end{equation}
\begin{equation}
\mathcal{I}_{ijkl}:=\{(i,j,l,k)|i\neq j;l\neq k;i\neq l\hspace{1mm} or \hspace{1mm} j \neq k\}
\end{equation}
\begin{equation}
\mathcal{I}_{ij}:=\{(i,j)|i\neq j\},\;\;
\omega_{ij}:=\{(x,y)|x\in\Delta_{i};y\in\Delta_{j}\}
\end{equation}
\begin{equation}
\Omega_{ijkl}:=\{(x,y,w,v)|x\in\Delta_{i};y\in\Delta_{j};w\in\Delta_{l};v\in\Delta_{k}\}
\end{equation}
\end{Co}
The proof is a direct application of Theorem \ref{eqn:themain}.
\subsection{Effective equilibration for continuous variables}
\;\;\;In order to develop estimates for effective equilibration, we will employ the same approach taken in\cite{short}, albeit with adaptations made to accommodate for the case of continuous variables. For this, let us first adopt the following definition used in \cite{short}.
\begin{definition}[\textbf{Distinguishability with respect to a set of POVM}]
\begin{equation}
D_{\mathscr{M}}\big(\boldsymbol{\hat{\rho}}_{1}, \boldsymbol{\hat{\rho}}_{2}\big) := \max_{M_{l}\in\mathscr{M}}\frac{1}{2}\sum_{r}\big|Tr\big\{\mathbf{\hat{M}}_{lr}\boldsymbol{\hat{\rho}}_{1}\big\}-Tr\big\{\mathbf{\hat{M}}_{lr}\boldsymbol{\hat{\rho}}_{2}\big\}\big|
\end{equation}
$M_{l}$ represents the POVM \cite{Nielsen} $\big\{\mathbf{\hat{M}}_{lr}\big\}_{r}$, while $\mathscr{M}$ represents the set of POVMs $\big\{M_{l}\big\}_{l}$.
\end{definition}
With the latter in mind, we may now define $T$-time effective equilibration. 
\begin{definition}[$T$\textbf{-time effective equilibration of a quantum state} $\boldsymbol{\hat{\rho}}_{t}$]
We say that a quantum state $\boldsymbol{\hat{\rho}}_{t}$ is $T-$time effective equilibrated with respect to a family of POVM measurements $\mathscr{M}$ with tolerance $\delta>0$ when 
\begin{equation}
\bigg\langle\!\!\!\bigg\langle D_{\mathscr{M}}\Big(\boldsymbol{\hat{\rho}}_{t}, \big\langle\!\!\big\langle\boldsymbol{\hat{\rho}}_{t}\big\rangle\!\!\big\rangle_{T}\Big)\bigg\rangle\!\!\!\bigg\rangle_{T}<\delta
\end{equation}
\end{definition}
In \cite{short}, the set $\mathscr{M}$ is assumed to be the set of realistic quantum measurements (POVM); these sets are finite in such a setting owing to the finite dimensionality of the Hilbert spaces in question. In this work, rather than getting it for free, we must assume that all of the POVMs $M_{l}$ are finite-dimensional. Physically this makes sense, since measurement apparatuses are invariably calibrated to detect a finite range of outcomes with resolution limitations. With the latter assumption in mind, let us return to the setting pertaining to the hypothesis of Theorem \ref{eqn:themain}. We will now make the definition below.
\begin{equation}
\frac{1}{3}\mathcal{B}(T,\Delta):=\Big(\sum_{\mathcal{J}_{ij}}\sup_{\omega_{ij}}F(T,\Delta,x,y)\Big)^{2} + \mathscr{F}(T,\Delta) +
\sum_{i}\beta_{i}^{2}+
\sum_{\mathcal{I}_{ijlk}}\sup_{\Omega_{ijlk}}F(T,\Delta,x-v, y-w)
\end{equation}
Furthermore, consider the following inequalities. 
\begin{equation}
\bigg\langle\!\!\!\bigg\langle D_{\mathscr{M}}\Big(\boldsymbol{\hat{\rho}}_{t}, \big\langle\!\!\big\langle\boldsymbol{\hat{\rho}}_{t}\big\rangle\!\!\big\rangle_{T}\Big)\bigg\rangle\!\!\!\bigg\rangle_{T} = \bigg\langle\!\!\!\bigg\langle \max_{M_{l}(t)\in\mathscr{M}}\frac{1}{2}\sum_{r}\big|Tr\big\{\mathbf{\hat{M}}_{lr}\boldsymbol{\hat{\rho}}_{t}\big\}-Tr\big\{\mathbf{\hat{M}}_{lr}\big\langle\!\!\big\langle\boldsymbol{\hat{\rho}}_{t}\big\rangle\!\!\big\rangle_{T}\big\}\big|\bigg\rangle\!\!\!\bigg\rangle_{T} \leq
\end{equation}
\begin{equation}
\frac{1}{2}\sum_{M_{l}\in\mathscr{M}}\sum_{r}\bigg\langle\!\!\!\bigg\langle \big|Tr\big\{\mathbf{\hat{M}}_{lr}\boldsymbol{\hat{\rho}}_{t}\big\}-Tr\big\{\mathbf{\hat{M}}_{lr}\big\langle\!\!\big\langle\boldsymbol{\hat{\rho}}_{t}\big\rangle\!\!\big\rangle_{T}\big\}\big|\bigg\rangle\!\!\!\bigg\rangle_{T}\leq
\end{equation}
\begin{equation}
\frac{1}{2}\sum_{M_{l}\in\mathscr{M}}\sum_{r}\sqrt{\bigg\langle\!\!\!\bigg\langle \big|Tr\big\{\mathbf{\hat{M}}_{lr}\boldsymbol{\hat{\rho}}_{t}\big\}-Tr\big\{\mathbf{\hat{M}}_{lr}\big\langle\!\!\big\langle\boldsymbol{\hat{\rho}}_{t}\big\rangle\!\!\big\rangle_{T}\big\}\big|^{2}\bigg\rangle\!\!\!\bigg\rangle_{T}} =
\end{equation}
\begin{equation}
\frac{1}{2}\sum_{M_{l}\in\mathscr{M}}\sum_{r}\sigma_{\mathbf{\hat{M}}_{lr}}(T) \leq
\end{equation}
\begin{equation}
Q(\mathscr{M})\sqrt{3\mathcal{B}(T,\Delta)} 
\end{equation}
where $Q(\mathscr{M})$ is half the total number of outcomes for all measurements in $\mathscr{M}$. This bound allows us to bound T-time effective equilibration for the case of continuous variables; we present it as a corollary below.  
\begin{Co}[$T$-\textbf{time effective equilibration estimates for continuous variables} ]
\label{eqn:yoyo}
Following the setup of Theorem \ref{eqn:themain}.
\begin{equation}
\bigg\langle\!\!\!\bigg\langle D_{\mathscr{M}}\Big(\boldsymbol{\hat{\rho}}_{t},\big\langle\!\!\big\langle\boldsymbol{\hat{\rho}}_{t}\big\rangle\!\!\big\rangle_{T}\Big)\bigg\rangle\!\!\!\bigg\rangle_{T} \leq Q(\mathscr{M})\sqrt{3\mathcal{B}(T,\Delta)} 
\end{equation}
where $Q(\mathscr{M})$ is half the total number of outcomes for all of the POVMS in $\mathscr{M}$ combined, and $\mathcal{B}(T,\Delta)$ is the bound in Theorem \ref{eqn:themain}.
\end{Co}
In the following section, we present a simple example of the tools developed in this section.

\end{section}

\begin{section}{Toy example}
\label{eqn:sec6}
\;\;\;Consider the setting described by Theorem \ref{eqn:themain2}, where we now define the function $f(x)=x^{2}$, for $x\in\Omega\subset \mathbb{R}^{+}_{0}$; zero elsewhere. Furthermore, assume that $\psi(p)= \frac{1}{N}\sum_{i=1}^{N}\chi_{\Delta_{i}}(p)$ and $\Delta_{i}\subset \Omega$, whence $|\psi(p)|^{2} = \frac{1}{N^{2}}\sum_{i=1}\chi_{\Delta_{i}}(p)$. Now, notice that 
\begin{equation}
\beta_{i}:=\int_{\Delta_{i}}|\psi(p)|^{2}dp= \int_{\Delta_{i}}\frac{1}{N^{2}}\bigg(\sum_{i=1}^{N}\chi_{\Delta_{i}}(p)\bigg)dp = \frac{\Delta}{N^{2}}
\end{equation}
Furthermore,
\begin{equation}
\int_{\Delta_{i}}e^{-i\tau p^{2}}|\psi(p)|^{2}dp = \frac{1}{N^{2}}\int_{\Delta_{i}}e^{-i\tau p^{2}}\bigg(\sum_{i=1}^{N}\chi_{\Delta_{i}}(p)\bigg)dp = \frac{1}{N^{2}}\int_{\Delta_{i}} e^{-i\tau p^{2}}dp =
\end{equation}
\begin{equation}
\frac{\Delta}{N^{2}}\int_{0}^{1} e^{-i\tau (\Delta x + a_{i})^{2}}dx
\end{equation}
Hence, for this example, letting $\tau_{*}$ and $a_{*}$ be respectively the value in $[-T, T]$ and $\{a_{i}\}_{i=1}^{N}$, where $a_{i}$ are the left-end points of $\Delta_{i}$, that maximizes the right hand side of (\ref{eqn:coke}) (where there is only one index i since we are dealing with a pure state at $t=0$).
\begin{equation}
4\max_{i}\sup_{\tau\in[-T,T]}\Bigg(1-\beta_{i}^{-2}\Big|\int_{\Delta_{i}}e^{-i\tau p^{2}}|\psi(p)|^{2}dp\Big|^{2}\Bigg) = 4\Bigg(1-\frac{N^{4}}{\Delta^{2}}\bigg|\frac{\Delta}{N^{2}}\int_{0}^{1} e^{-i\tau_{*} (\Delta x + a_{*})^{2}}dx\bigg|^{2}\Bigg) = 
\end{equation}
\begin{equation}
4\Bigg(1-\big|g(\Delta \tau_{*})\big|^{2}\Bigg)    
\end{equation}
where $g(x): = \int_{0}^{1} e^{-i\tau_{*} (\Delta x + a_{*})^{2}}dx$. Furthermore, we will now bound $F(T, \Delta, x, y)$. Using the definition of the $\textbf{sinc}(x)$ function, it can  easily be shown that
\begin{equation}
F(T, \Delta, x, y) \leq \frac{4+2\Delta T}{|x-y|T}+\frac{2\Delta T}{|x-y|^{2}T^{2}}
\end{equation}
which implies the following. 
\begin{equation}
\sum_{\mathcal{J}_{ij}}\sup_{\omega_{ij}}F\big(T,\Delta,x^{2},y^{2}) \leq \sum_{i\neq j}\sup_{(x,y)\in \Delta_{i}\times \Delta_{j}}\bigg(\frac{4+2\Delta T}{|x-y|T}+\frac{2\Delta T}{|x-y|^{2}T^{2}}\bigg) \leq 
\end{equation}
\begin{equation}
N^{2}\bigg(\frac{4+2\Delta T}{DT}+\frac{2\Delta T}{D^{2}T^{2}}\bigg)
\end{equation}
Where we define $D:= \min_{i\neq j}dist\big(f(\Delta_{i}),f(\Delta_{j})\big)$. Similarly,
\begin{equation}
\sum_{\mathcal{I}_{ijlk}}\sup_{\Omega_{ijlk}}F(T,\Delta,x^{2}-v^{2},w^{2}-y^{2})\leq
\end{equation}
\begin{equation}
\leq \sum_{\mathcal{I}_{ijlk}}\sup_{\Omega_{ijlk}}\bigg(\frac{4+2\Delta T}{|x^{2}-v^{2}+w^{2}-y^{2}|T}+\frac{2\Delta T}{|x^{2}-v^{2}+w^{2}-y^{2}|^{2}T^{2}}\bigg) \leq
\end{equation}
\begin{equation}
N^{4}\bigg(\frac{4+2\Delta T}{DT}+\frac{2\Delta T}{D^{2}T^{2}}\bigg) = \frac{2N^{4}}{DT}\bigg(2+\Delta T\Big(1+\frac{1}{DT}\Big)\bigg)
\end{equation}
The latter is so, since either $x^{2}-v^{2}$ or $w^{2}-y^{2}$ may be equal to zero but not both. In conclusion, we have the following result.
\begin{equation}
\Big\langle\!\!\!\Big\langle\big|Tr\big\{\mathbf{\hat{A}}\boldsymbol{\hat{\rho}}_{t}\big\}- Tr\big\{\mathbf{\hat{A}}\big\langle\!\!\big\langle\boldsymbol{\hat{\rho}}_{t}\big\rangle\!\!\big\rangle_{T}\big\}\big|^{2} \Big\rangle\!\!\!\Big\rangle_{T}\leq 
\end{equation}
\begin{equation}
3N^{4}H(\Delta, D, T)\Big( 1 + H(\Delta, D, T)\Big) + 12\Big(1-\Big|g(\Delta \tau_{*})\Big|^{2}\Big)+\frac{3\Delta}{N^{2}}
\end{equation}
where we have defined $H(\Delta, D, T):= \frac{2}{DT}(2+\Delta T(1+\frac{1}{DT}))$. Indeed, for the latter to be small, $\frac{\Delta}{N^{2}}$ needs to be small, which implies that $\Delta << N^{2}$. Secondly, we must have $|g(\Delta \tau_{*})|^{2}\approx 1$, which means that we must have $\Delta^{2}\tau_{*} \approx 0$; indirectly requiring that $\Delta^{2} T\approx 0$. Finally, for the term $\frac{2N^{4}}{DT}(2+\Delta T(1+\frac{1}{DT}))$ to be small it is sufficient that $N^{4}<<DT$ and $\Delta < D$.\\

In the latter, we have developed the estimates necessary to control $\sigma_{\mathbf{\hat{A}}}^{2}(T)$ for a given desired tolerance. If we now would like $T$-time effective equilibration to take place with a tolerance of $\delta$; i.e. $\langle\!\langle D_{\mathscr{M}}(\boldsymbol{\hat{\rho}}_{t}, \langle\boldsymbol{\hat{\rho}}_{t}\rangle_{T})\rangle\!\rangle_{T} \leq \delta$, then by applying Corollary \ref{eqn:yoyo} we need only require that $\sigma_{\mathbf{\hat{A}}}^{2}(T)\leq \frac{\delta^{2}}{3Q^{2}(\mathscr{M})\mathcal{B}(T,\Delta)}$.
\end{section}

\begin{section}{Conclusion}
Theorems \ref{eqn:themain}, Corollary \ref{eqn:themain2}, and Corollary \ref{eqn:yoyo} are continuous variable analogues to the estimates on equilibration-on-average and effective equilibration derived in \cite{short}. The term $\sum_{i}\beta_{i}^{2}$, appearing in both Theorem \ref{eqn:themain} and Corollary \ref{eqn:themain2}, is the analogue to the inverse effective dimension $d_{eff}^{-1}$ that shows up in the estimates for the finite-dimensional case, see Theorem \ref{eqn:themainresultriemann}. However, the other terms, independent of the effective dimension for continuous variables $\sum_{i}\beta_{i}^{2}$ , appearing in the estimates of Theorem \ref{eqn:themain} Corollary \ref{eqn:themain2}, and Corollary \ref{eqn:yoyo} have no analogue in the finite-dimensional case. The estimates provided in this work illustrate sufficient conditions for a quantum state to be equilibrated on average and effectively equilibrate to some desired tolerance of error. These estimates depend on the physical parameters/initial conditions as $\Delta$, $D$, $T$, $N$, and $|\psi\rangle$ respectively characterizing the resolution limit of the spectrum of $\mathbf{\hat{H}}$, the distance between macroscopic coherences, the finite evolution time, the size of the partitioning of the time-evolved state induced by the resolution limit parameter $\Delta$ and the compactness of $\sigma(\mathbf{\hat{H}})$, and the initial state of the system under study.  The example provided in Section \ref{eqn:sec6} shows an instance of this. There, after some calculation, we can identify the relevant relationships between parameters and initial conditions  $\Delta$, $D$ , $T$, $N$ $|\psi\rangle$. Namely, for this example, so long as $N<<DT$, $\Delta<<N^{2}$, $\Delta< D$ and $\Delta T\approx 0$ we should achieve equilibration-on-average and approximate effective equilibration. Here, parameters may be tampered with in such a way as to yield the desired level of effective equilibration by pre-selecting an error tolerance. Looking to the future, it would be interesting to find out if the restrictions imposed by requiring the bounds of Theorem \ref{eqn:themain} and Corollary \ref{eqn:themain2}to be small might also be necessary conditions for equilibration in a generic sense. Also, for the same theorems, we have assumed a pure initial state; in the future we will further look into adapting these theorems to the case of a general state.
\end{section}

\vspace{1cm}

\noindent \textbf{Funding:} This research was funded by the grant number COMCUANTICA/007 from the Generalitat Valenciana, Spain and by the grant number INDI24/17 from the Universidad CEU Cardenal Herrera, Spain.

\newpage
\appendix 
\addcontentsline{toc}{section}{Appendices}
\section{Proof of Lemma \ref{eqn:theoremkupsch} }
\label{eqn:appa}
\begin{proof}
\vspace{4mm}
CASE 1) \\

We assume that $\sup \big\{\boldsymbol{\hat{\rho}}_{0}\big\}\cap \sup \big\{\mathbf{\hat{P}}(\Delta_{j})\big\} = 0$. Then, 
\begin{equation}
\Big\|\mathbf{\hat{P}}(\Delta_{i})\boldsymbol{\hat{\rho}}_{t}\mathbf{\hat{P}}(\Delta_{j})\Big\|_{1} = \bigg\|\mathbf{\hat{P}}(\Delta_{i})\Big(\int\int  \Gamma(t,x,y)\mathbf{\hat{E}}(dx)\boldsymbol{\hat{\rho}}_{0}\mathbf{\hat{E}}(dy)\Big)\mathbf{\hat{P}}(\Delta_{j})\bigg\|_{1}
\end{equation}
\begin{equation}
\bigg\|\Big(\int_{\Delta_{i}}\int_{\Delta_{j}}  \Gamma(t,x,y)\mathbf{\hat{E}}(dx)\boldsymbol{\hat{\rho}}_{0}\mathbf{\hat{E}}(dy)\Big)\bigg\|_{1} = \bigg\|\Big(\int\int  \Gamma(t,x,y)\mathbf{\hat{E}}(dx)\mathbf{\hat{P}}(\Delta_{i})\boldsymbol{\hat{\rho}}_{0}\mathbf{\hat{P}}(\Delta_{j})\mathbf{\hat{E}}(dy)\Big)\bigg\|_{1} =
\end{equation}
\begin{equation}
\bigg\|\Big(\int\int  \Gamma(t,x,y)\mathbf{\hat{E}}(dx)\mathbf{\hat{P}}(\Delta_{i})\big\{0\big\}\mathbf{\hat{E}}(dy)\Big)\bigg\|_{1}  = 0 
\end{equation}
\vspace{4mm}
CASE 2)
Now we assume that 
\begin{equation}
\sup \big\{\boldsymbol{\hat{\rho}}_{0}\big\}\cap \sup \big\{\mathbf{\hat{P}}(\Delta_{i})\big\} \neq 0
\end{equation}
and
\begin{equation}
\sup \big\{\boldsymbol{\hat{\rho}}_{0}\big\}\cap \sup \big\{\mathbf{\hat{P}}(\Delta_{j})\big\} \neq 0
\end{equation}
Let us begin by considering the operator 
 \begin{equation}
\mathbf{\hat{T}}_{t}(y):= \int_{\Delta_{i}} \Gamma(t,x,y)\mathbf{\hat{E}}(dx)
 \end{equation}
The set $\{\mathbf{\hat{T}}_{t}(y)\}_{t\in\mathbb{R}_{+}}$ is a family of operators, differentiable with respect to $y$,  satisfying the operator norm estimate
\begin{equation} 
 \big\|\mathbf{\hat{T}}_{t}(y)\big\|\leq \sup_{x\in\Delta_{i}}|\Gamma(t,x,y)|
 \end{equation}
\begin{proof}
 \begin{equation}
\big\|\mathbf{\hat{T}}_{t}(y)\big\|^{2} = \sup_{\||\psi\rangle\|=1}\big\|\mathbf{\hat{T}}_{t}(y)\big|\psi\big\rangle\big\|^{2}=
\end{equation}
\begin{equation}
\sup_{\||\psi\rangle\|=1}\int_{\Delta_{i}}\int_{\Delta_{i}} \Gamma(t,x^\prime,y)^{*}\Gamma(t,x,y)\big\langle \psi\big|\mathbf{\hat{E}}(dx^\prime)\mathbf{\hat{E}}(dx)\big|\psi\big\rangle dx^\prime dx=
\end{equation}
\begin{equation}
\sup_{\||\psi\rangle\|=1}\int_{\Delta_{i}} |\Gamma(t,x,y)|^{2}\big\langle \psi\big|\mathbf{\hat{E}}(dx)|\psi\big\rangle dx\leq 
\end{equation}
\begin{equation}
\sup_{x\in \Delta_{i}}|\Gamma(t,x,y)|^{2}\sup_{\||\psi\rangle\|=1}\int_{\Delta_{i}}\mu_{\psi}(dx) \leq \sup_{x\in \Delta_{i}}|\Gamma(t,x,y)|^{2}
\end{equation}
\end{proof}
\vspace{5mm}
In a similar way, we may bound the operator norm of $\mathbf{\hat{T}}^{\prime}_{t}(y):=\int_{\Delta_{i}}\Gamma^{'}(t,x,y)\mathbf{\hat{E}}(dx)$
where $\Gamma^{'}(t,x,y):=\partial_{y}\Gamma(t,x,y)$. Namely, 
\begin{equation}
\big\|\mathbf{\hat{T}}^{\prime}_{t}(y)\big\|\leq \sup_{x\in \Delta_{i}}|\Gamma^{'}(t,x,y)|
\end{equation}
Furthermore, define $\mathbf{\hat{J}}_{t}(y):= \mathbf{\hat{T}}_{t}(y)\boldsymbol{\hat{\rho}}_{0}$ and $\mathbf{\hat{J}}_{t}^{'}(y):=\mathbf{\hat{T}}_{t}^{'}(y)\boldsymbol{\hat{\rho}}_{0} $. Using the above operator norm estimates and the inequality $\|\mathbf{\hat{A}}\mathbf{\hat{C}}\|_{1}\leq \|\mathbf{\hat{A}}\|\|\mathbf{\hat{C}}\|_{1}$ \cite{simon3} we obtain
\begin{equation}
\big\|\mathbf{\hat{J}}_{t}(y)\big\|_{1}\leq \sup_{x\in \Delta_{i,t}}|\Gamma(t,x,y)|\big\|\boldsymbol{\hat{\rho}}_{0}\big\|_{1} = \sup_{x\in \Delta_{i,t}}|\Gamma(t,x,y)|
\end{equation}
 and 
\begin{equation}
\big\|\mathbf{\hat{J}}_{t}^{'}(y)\big\|_{1}\leq \sup_{x\in \Delta_{i,t}}|\Gamma^{'}(t,x,y)|\big\| \boldsymbol{\hat{\rho}}_{0} \big\|_{1}=  \sup_{x\in \Delta_{i,t}}|\Gamma^{'}(t,x,y)|.
\end{equation}
We now elucidate on the relationship between the operator $\mathbf{\hat{T}}_{t}^{\prime}(y)$ and the derivative $\partial_{y}\langle \psi|\mathbf{\hat{T}}_{t}(y)|\phi\rangle$.
\begin{equation}
\partial_{y}\big\langle \psi\big|\mathbf{\hat{T}}_{t}(y)\big|\phi\big\rangle = \partial_{y}\int_{\Delta_{i}}\Gamma(t,x,y)\big\langle \psi\big|\mathbf{\hat{E}}(dx)\big|\phi\big\rangle  
\end{equation} 
Given that $\Gamma(t,x,y)$ is smooth with respect to $y$, we may swap the order of the integral and the derivative. 
\begin{equation}
\partial_{y}\int_{\Delta_{i}}\Gamma(t,x,y)\big\langle \psi\big|\mathbf{\hat{E}}(dx)\big|\phi\big\rangle dx = \int_{\Delta_{i}}\partial_{y}\Gamma(t,x,y)\big\langle \psi\big|\mathbf{\hat{E}}(dx)\big|\phi\big\rangle = 
\end{equation}
\begin{equation}
\int_{\Delta_{i}}\Gamma^{'}(t,x,y)\big\langle \psi\big|\mathbf{\hat{E}}(dx)\big|\phi\big\rangle  = \big\langle \psi\big|\Big(\int_{\Delta_{i}}\Gamma^{'}(t,x,y)\mathbf{\hat{E}}(dx)\Big)\big|\phi\big\rangle = \big\langle \psi\big|\boldsymbol{\hat{T}}_{t}^{\prime}(y)\big|\phi\big\rangle
\end{equation}
We therefore conclude that
\begin{equation}
 \partial_{y}\big\langle \psi\big|\mathbf{\hat{T}}_{t}(y)\big|\phi\big\rangle =   \big\langle \psi\big|\mathbf{\hat{T}}_{t}^{\prime}(y)\big|\phi \big\rangle
\end{equation}
Now, for the interval $\Delta_{j}$, we have $\int_{\Delta_{j}}\mathbf{\hat{J}}_{t}(y)\mathbf{\hat{E}}(dy) = \mathbf{\hat{P}}(\Delta_{i})\boldsymbol{\hat{\rho}}_{t}\mathbf{\hat{P}}(\Delta_{j})$. Let us further specify $\Delta_{j}:= [a_{j},b_{j}]$. We will show that the following identity holds
\begin{equation}
\label{eqn:IBPFS}
\int_{\Delta_{j}}\mathbf{\hat{J}}_{t}(y)\mathbf{\hat{E}}(dy)= \mathbf{\hat{J}}_{t}(b_{j})\mathbf{\hat{P}}((-\infty,b_{j}])-\mathbf{\hat{J}}_{t}(a_{j})\mathbf{\hat{P}}((-\infty,a_{j}]) - \int_{\Delta_{j}}\mathbf{\hat{J}}_{t}^\prime(y)\mathbf{\hat{P}}((-\infty,y])dy
\end{equation}
\begin{proof}
The state $\boldsymbol{\hat{\rho}}_{0}$ may be generally written as a mixture $\boldsymbol{\hat{\rho}}_{0}= \sum_{n}p_{n}\big|\xi_{0}^{n}\big\rangle\big\langle\xi_{0}^{n}\big|$. Now, for arbitrary $\big|\psi\big\rangle$ and $\big|\phi\big\rangle$  
\begin{equation}
\label{eqn:abouttodointbypart}
\big\langle\psi\big|\int_{\Delta_{j}}\mathbf{\hat{J}}_{t}(y)\mathbf{\hat{E}}(dy)\big|\phi\big\rangle =\int_{\Delta_{j}}\big\langle\psi\big|\mathbf{\hat{J}}_{t}(y)\mathbf{\hat{E}}(dy)\big|\phi\big\rangle
\end{equation}
By the definition of $\mathbf{\hat{J}}_{t}(y)$ one may write
\begin{equation}
\big\langle\psi\big|\mathbf{\hat{J}}_{t}(y)=\big\langle\psi\big|\mathbf{\hat{T}}_{t}(y)\boldsymbol{\hat{\rho}}_{0} = \big\langle\psi\big|\mathbf{\hat{T}}_{t}(y)\Big(\sum_{n}p_{n}\big|\xi_{0}^{n}\big\rangle\big\langle\xi_{0}^{n}\big|\Big)
\end{equation}
The right-hand side of $(\ref{eqn:abouttodointbypart})$ therefore equals
\begin{equation}
\sum_{n}p_{n}\int_{\Delta_{j}}\big\langle\psi\big|\mathbf{\hat{T}}_{t}(y)\big|\xi^{n}_{0}\big\rangle\big\langle \xi^{n}_{0}\big|\mathbf{\hat{E}}(dy)\big|\phi\big\rangle 
\end{equation}
Furthermore,
\begin{equation}
\sum_{n}p_{n}\int_{\Delta_{j}}\big\langle\psi\big|\mathbf{\hat{T}}_{t}(y)\big|\xi^{n}_{0}\big\rangle\big\langle \xi^{n}_{0}\big|\mathbf{\hat{E}}(dy)\big|\phi\big\rangle = 
\end{equation}
\begin{equation}
\sum_{n}p_{n}\bigg[\big\langle\psi|\mathbf{\hat{T}}_{t}(y)\big|\xi^{n}_{0}\big\rangle\big\langle\xi^{n}_{0}\big|\mathbf{\hat{E}}(y^\prime)\big|\phi\big\rangle \bigg]\Bigg|_{a_{j}}^{b_{j}}- \sum_{n}p_{n}\int_{\Delta_{j}}\big\langle \xi^{n}_{0}\big|\mathbf{\hat{E}}(y^\prime)\big|\phi\big\rangle  d\Big(\big\langle\psi\big|\mathbf{\hat{T}}_{t}(y)\big|\xi^{n}_{0}\big\rangle\Big) = \end{equation}
\begin{equation}
\sum_{n}p_{n}\bigg[\big\langle\psi|\mathbf{\hat{T}}_{t}(y)\big|\xi^{n}_{0}\big\rangle\big\langle\xi^{n}_{0}\big|\int_{-\infty}^{y}\mathbf{\hat{E}}(dy^\prime)\big|\phi\big\rangle \bigg]\Bigg|_{a_{j}}^{b_{j}}- \sum_{n}p_{n}\int_{\Delta_{j}}\big\langle \xi^{n}_{0}\big|\int_{-\infty}^{y}\mathbf{\hat{E}}(dy^\prime)\big|\phi\big\rangle  d\Big(\big\langle\psi\big|\mathbf{\hat{T}}_{t}(y)\big|\xi^{n}_{0}\big\rangle\Big) = \end{equation}
\begin{equation}
\sum_{n}p_{n}\bigg[\big\langle\psi\big|\mathbf{\hat{T}}_{t}(y)\big|\xi^{n}_{0}\big\rangle\big\langle\xi^{n}_{0}\big|\mathbf{\hat{P}}\big((-\infty,y]\big)\big|\phi\big\rangle \bigg]\Bigg|_{a_{j}}^{b_{j}}- \sum_{n}p_{n}\int_{\Delta_{j}}\big\langle \xi^{n}_{0}\big|\mathbf{\hat{P}}\big((-\infty,y]\big)|\phi\big\rangle\Big(\big\langle\psi\big|\mathbf{\hat{T}}_{t}(y)\big|\xi^{n}_{0}\big\rangle\Big)^\prime dy =     
\end{equation}
\begin{equation}
\sum_{n}p_{n}\bigg[\big\langle\psi\big|\mathbf{\hat{T}}_{t}(y)\big|\xi^{n}_{0}\big\rangle\big\langle\xi^{n}_{0}\big|\mathbf{\hat{P}}((-\infty,y])\big|\phi\big\rangle \bigg]\Bigg|_{a_{j}}^{b_{j}}- \sum_{n}p_{n}\int_{\Delta_{j}}\Big(\big\langle\psi\big|\mathbf{\hat{T}}_{t}(y)\big|\xi^{n}_{0}\big\rangle\Big)^\prime\big\langle \xi^{n}_{0}\big|\mathbf{\hat{P}}\big((-\infty,y]\big)\big|\phi\big\rangle dy =  \end{equation}
\begin{equation}
\sum_{n}p_{n}\big\langle\psi\big|\bigg[\mathbf{\hat{T}}_{t}(y)\big|\xi^{n}_{0}\big\rangle\big\langle\xi^{n}_{0}\big|\mathbf{\hat{P}}\big((-\infty,y]\big)\bigg]\Bigg|_{a_{j}}^{b_{j}}\big|\phi\big\rangle - \sum_{n}p_{n}\big\langle\psi\big|\bigg(\int_{\Delta_{j}}\mathbf{\hat{T}}^{\prime}_{t}(y)\big|\xi^{n}_{0}\big\rangle\big\langle \xi^{n}_{0}\big|\mathbf{\hat{P}}\big((-\infty,y]\big) dy \bigg)\big|\phi\big\rangle =    
\end{equation}
\begin{equation}
\big\langle\psi\big|\bigg[\mathbf{\hat{T}}_{t}(y)\boldsymbol{\hat{\rho}}_{0}\mathbf{\hat{P}}\big((-\infty,y]\big)\bigg]\Bigg|_{a_{j}}^{b_{j}}\big|\phi\big\rangle - \big\langle\psi\big|\bigg(\int_{\Delta_{j}}\mathbf{\hat{T}}^{\prime}_{t}(y)\boldsymbol{\hat{\rho}}_{0}\mathbf{\hat{P}}\big((-\infty,y]\big) dy \bigg)\big|\phi\big\rangle =    
\end{equation}
\begin{equation}
\big\langle\psi\big|\bigg(\mathbf{\hat{J}}_{t}(y)\mathbf{\hat{P}}\big((-\infty,y]\big)\Bigg|_{a_{j}}^{b_{j}}- \int_{\Delta_{j}}\mathbf{\hat{J}}^{\prime}_{t}(y) \mathbf{\hat{P}}\big((-\infty,y]\big) dy\bigg)\big|\phi\big\rangle  =  
\end{equation}
\begin{equation}
\big\langle\psi\big|\bigg(\mathbf{\hat{J}}_{t}(b_{j})\mathbf{\hat{P}}\big((-\infty,b_{j}]\big)-\hat{J}_{t}(a_{j})\mathbf{\hat{P}}\big((-\infty,a_{j}]\big)- \int_{\Delta_{j}}\mathbf{\hat{J}}_{t}^{\prime}(y) \mathbf{\hat{P}}\big((-\infty,y]) dy\bigg)\big|\phi\big\rangle   
\end{equation}
and so 
\begin{equation}
\int_{\Delta_{j}}\mathbf{\hat{J}}_{t}(y)\mathbf{\hat{E}}(dy)= \mathbf{\hat{J}}_{t}(b_{j})\mathbf{\hat{P}}\big((-\infty,b_{j}]\big)-\mathbf{\hat{J}}_{t}(a_{j})\mathbf{\hat{P}}\big((-\infty,a_{j}]\big) - \int_{\Delta_{j}}\mathbf{\hat{J}}_{t}^\prime(y)\mathbf{\hat{P}}\big((-\infty,y]\big)dy
\end{equation}
\end{proof}
In consequence to the latter. 
\begin{equation}
\big\|\mathbf{\hat{P}}\big(\Delta_{i}\big)\boldsymbol{\hat{\rho}}_{t}\mathbf{\hat{P}}\big(\Delta_{j}\big)\big\|_{1} =\bigg\|\int_{\Delta_{j}}\mathbf{\hat{J}}_{t}(y)\mathbf{\hat{P}}(dy)\bigg\|_{1}=  
\end{equation}
\begin{equation}
\bigg\|\mathbf{\hat{J}}_{t}(b_{j})\mathbf{\hat{P}}\big((-\infty,b_{j}]\big)-\mathbf{\hat{J}}_{t}(a_{j})\mathbf{\hat{P}}\big((-\infty,a_{j}]\big)- \int_{\Delta_{j}}\mathbf{\hat{J}}_{t}^{\prime}(y) \mathbf{\hat{P}}\big((-\infty,y]\big) dy\bigg\|_{1}\leq
\end{equation}
\begin{equation}
\big\|\mathbf{\hat{J}}_{t}(b_{j})\mathbf{\hat{P}}\big((-\infty,b_{j}]\big)\big\|_{1}+\big\|\mathbf{\hat{J}}_{t}(a_{j})\mathbf{\hat{P}}\big((-\infty,a_{j}]\big)\big\|_{1}+\bigg\| \int_{\Delta_{j}}\mathbf{\hat{J}}_{t}^{\prime}(y) \mathbf{\hat{P}}\big((-\infty,y]\big) dy\bigg\|_{1}\leq
\end{equation}
\begin{equation}
\big\|\mathbf{\hat{J}}_{t}(b_{j})\big\|_{1}\big\|\mathbf{\hat{P}}\big((-\infty,b_{j}]\big)\big\|+\big\|\mathbf{\hat{J}}_{t}(a_{j})\big\|_{1}\big\|\mathbf{\hat{P}}\big((-\infty,a_{j}]\big)
\big\|+\int_{\Delta_{j}}\big\|\mathbf{\hat{J}}_{t}(y)^\prime \mathbf{\hat{P}}\big((-\infty,y]\big)\big\|_{1} dy\leq
\end{equation}
\begin{equation}
\big\|\mathbf{\hat{J}}_{t}(b_{j})\big\|_{1}+\big\|\mathbf{\hat{J}}_{t}(a_{j})\big\|_{1}+\int_{\Delta_{j}}\big\|\mathbf{\hat{J}}_{t}^\prime(y)\big\|_{1}\big\| \mathbf{\hat{P}}\big((-\infty,y]\big)\big\|dy =
\end{equation}
\begin{equation}
\big\|\mathbf{\hat{J}}_{t}(b_{j})\big\|_{1}+\big\|\mathbf{\hat{J}}_{t}(a_{j})\big\|_{1}+\int_{\Delta_{j}}\big\|\mathbf{\hat{J}}_{t}^\prime(y)\big\|_{1}dy \leq 
\end{equation}
\begin{equation}
\big\|\mathbf{\hat{J}}_{t}(b_{j})\big\|_{1}+\big\|\mathbf{\hat{J}}_{t}(a_{j})\big\|_{1}+|\Delta_{j}|\sup_{y\in\Delta_{j}}\big\|\mathbf{\hat{J}}_{t}^\prime(y)\big\|_{1}dy \leq 
\end{equation}
\begin{equation}
\sup_{x\in \Delta_{i}}|\Gamma(t,x,b_{j})|+\sup_{x\in \Delta_{i}}|\Gamma(t,x,a_{j})|+ |\Delta_{j}|\sup_{\overset{x\in \Delta_{i}}{ y\in \Delta_{j}}}|\partial_{y}\Gamma(t,x,y)|\leq
\end{equation}
\begin{equation}
\sup_{\overset{(x,y)\in}{ \Delta_{i}\times \Delta_{j}}}\bigg(2|\Gamma(t,x,y)|+|\Delta_{j}||\Gamma^{'}(t,x,y)|\bigg)
\end{equation}
\end{proof}
\section{Proof of inequality (\ref{eqn:coke})}
\label{eqn:appb}
\begin{proof}
\begin{equation}
\mathscr{D}_{\mathbf{\hat{A}}}\big(T,\Delta\big):=\bigg\langle\!\!\!\bigg\langle\Big|Tr\Big\{\mathbf{\hat{A}}\sum_{i}\mathbf{\hat{P}}(\Delta_{i})\boldsymbol{\hat{\rho}}_{t} \mathbf{\hat{P}}(\Delta_{i})\Big\}- Tr\Big\{\mathbf{\hat{A}}\sum_{i}\mathbf{\hat{P}}(\Delta_{i})\big\langle\!\!\big\langle\boldsymbol{\hat{\rho}}_{t}\big\rangle\!\!\big\rangle_{T} \mathbf{\hat{P}}(\Delta_{i})\Big\}\Big|^{2}\bigg\rangle\!\!\!\bigg\rangle_{T} = 
\end{equation}
\begin{equation}
\bigg\langle\!\!\!\bigg\langle\Big|Tr\Big\{\mathbf{\hat{A}}\bigg(\sum_{i}\mathbf{\hat{P}}(\Delta_{i})\boldsymbol{\hat{\rho}}_{t} \mathbf{\hat{P}}(\Delta_{i})-\sum_{i}\mathbf{\hat{P}}(\Delta_{i})\big\langle\!\!\big\langle\boldsymbol{\hat{\rho}}_{t}\big\rangle\!\!\big\rangle_{T} \mathbf{\hat{P}}(\Delta_{i})\bigg)\Big\}\Big|^{2}\bigg\rangle\!\!\!\bigg\rangle_{T} \leq
\end{equation}
\begin{equation}
\bigg\langle\!\!\!\bigg\langle\Big\|\sum_{i}\mathbf{\hat{P}}(\Delta_{i})\boldsymbol{\hat{\rho}}_{t} \mathbf{\hat{P}}(\Delta_{i})-\sum_{i}\mathbf{\hat{P}}(\Delta_{i})\big\langle\!\!\big\langle\boldsymbol{\hat{\rho}}_{t}\big\rangle\!\!\big\rangle_{T} \mathbf{\hat{P}}(\Delta_{i})\Big\|_{1}^{2}\bigg\rangle\!\!\!\bigg\rangle_{T} \leq
\end{equation}
\begin{equation}
\bigg\langle\!\!\!\bigg\langle\bigg(\sum_{j}\alpha_{j}\Big\|\sum_{i}\mathbf{\hat{P}}(\Delta_{i})\big|\psi_{j}(t)\big\rangle\big\langle \psi_{j}(t)\big|\mathbf{\hat{P}}(\Delta_{i})-\sum_{i}\mathbf{\hat{P}}(\Delta_{i})\Big\langle\!\!\Big\langle\big|\psi_{j}(t)\big\rangle\big\langle \psi_{j}(t)\big|\Big\rangle\!\!\Big\rangle_{T} \mathbf{\hat{P}}(\Delta_{i})\Big\|_{1}\bigg)^{2}\bigg\rangle\!\!\!\bigg\rangle_{T} \leq
\end{equation}
\begin{equation}
\bigg\langle\!\!\!\bigg\langle\bigg(\sum_{j}\alpha_{j}\sum_{i}\Big\|\mathbf{\hat{P}}(\Delta_{i})\big|\psi_{j}(t)\big\rangle\big\langle \psi_{j}(t)\big|\mathbf{\hat{P}}(\Delta_{i})-\mathbf{\hat{P}}(\Delta_{i})\Big\langle\!\!\Big\langle\big|\psi_{j}(t)\big\rangle\big\langle \psi_{j}(t)\big|\Big\rangle\!\!\Big\rangle_{T} \mathbf{\hat{P}}(\Delta_{i})\Big\|_{1}\bigg)^{2}\bigg\rangle\!\!\!\bigg\rangle_{T} =
\end{equation}
\small
\begin{equation}
\bigg\langle\!\!\!\bigg\langle\bigg(\sum_{j}\alpha_{j}\sum_{i}\big\langle \psi_{j}(t)\big|\mathbf{\hat{P}}(\Delta_{i})\big|\psi_{j}(t)\big\rangle\Bigg\|\frac{\mathbf{\hat{P}}(\Delta_{i})\big|\psi_{j}(t)\big\rangle\big\langle \psi_{j}(t)\big|\mathbf{\hat{P}}(\Delta_{i})}{\big\langle \psi_{j}(t)\big|\mathbf{\hat{P}}(\Delta_{i})\big|\psi_{j}(t)\big\rangle}-\frac{\mathbf{\hat{P}}(\Delta_{i})\Big\langle\!\!\!\Big\langle\big|\psi_{j}(t)\big\rangle\big\langle \psi_{j}(t)\big|\Big\rangle\!\!\!\Big\rangle_{T} \mathbf{\hat{P}}(\Delta_{i})}{\big\langle \psi_{j}(t)\big|\mathbf{\hat{P}}(\Delta_{i})\big|\psi_{j}(t)\big\rangle}\Bigg\|_{1}\bigg)^{2}\bigg\rangle\!\!\!\bigg\rangle_{T} \leq
\end{equation}
\normalsize
\begin{equation}
\label{eqn:genius}
\bigg\langle\!\!\!\bigg\langle\bigg(\sum_{j}\alpha_{j}\sum_{i}\beta_{ij}(t)\sqrt{Tr^{2}\big\{ \mathbf{\hat{C}}_{ij}(t)+\mathbf{\hat{B}}_{ij}(T)\big\}-4F^{2}\big( \mathbf{\hat{C}}_{ij}(t), \mathbf{\hat{B}}_{ij}(T)\big)}\bigg)^{2}\bigg\rangle\!\!\!\bigg\rangle_{T} 
\end{equation}
where we have used the inequality, $\|\mathbf{\hat{A}}-\mathbf{\hat{B}}\|_{1}\leq\sqrt{Tr^{2}\big\{\mathbf{\hat{A}}+\mathbf{\hat{B}}\big\}-4F^{2}(\mathbf{\hat{A}},\mathbf{\hat{B}})}$ from \cite{genfid} \cite{genfid1}, defining the fidelity $F(\mathbf{\hat{A}},\mathbf{\hat{B}}) :=\|\mathbf{\hat{A}}^{1/2}\mathbf{\hat{B}}^{1/2}\|_{1}$, and making the following additional definitions
\begin{equation}
\beta_{ij}:=\big\langle \psi_{j}(t)\big|\mathbf{\hat{P}}(\Delta_{i})\big|\psi_{j}(t)\big\rangle = \big\langle \psi_{j}(0)\big|\mathbf{\hat{P}}(\Delta_{i})\big|\psi_{j}(0)\big\rangle
\end{equation}
\begin{equation}
\mathbf{\hat{C}}_{ij}(t):=\beta_{ij}^{-1}\mathbf{\hat{P}}(\Delta_{i})\big|\psi_{j}(t)\big\rangle\big\langle \psi_{j}(t)\big|\mathbf{\hat{P}}(\Delta_{i})
\end{equation}
\begin{equation}
\mathbf{\hat{B}}_{ij}(T):=\beta_{ij}^{-1}\mathbf{\hat{P}}(\Delta_{i})\Big\langle\!\!\Big\langle\big|\psi_{j}(t)\big\rangle\big\langle \psi_{j}(t)\big|\Big\rangle\!\!\Big\rangle_{T}\mathbf{\hat{P}}(\Delta_{i})
\end{equation}
Now,
\begin{equation}
(\ref{eqn:genius})\leq
\Bigg\langle\!\!\!\Bigg\langle\sum_{j}\alpha_{j}\sum_{i}\beta_{ij}\Big(Tr^{2}\big\{ \mathbf{\hat{C}}_{ij}(t)+\mathbf{\hat{B}}_{ij}(T)\big\}-4F^{2}\big( \mathbf{\hat{C}}_{ij}(t), \mathbf{\hat{B}}_{ij}(T)\big)\Big)\bigg\rangle\!\!\!\bigg\rangle_{T} 
\end{equation}
Now, notice that 
\begin{equation}
F^{2}\big( \mathbf{\hat{C}}_{ij}(t), \mathbf{\hat{B}}_{ij}(T)\big) = \Big\|\mathbf{\hat{C}}^{1/2}_{ij}(t)\mathbf{\hat{B}}^{1/2}_{ij}(T)\Big\|_{1}^{2} =
\end{equation}
\begin{equation}
Tr^{2}\bigg\{\sqrt{\sqrt{\mathbf{\hat{C}}_{ij}(t)}\mathbf{\hat{B}}_{ij}(T) \sqrt{\mathbf{\hat{C}}_{ij}(t)}}\Bigg\} = Tr^{2}\Bigg\{\sqrt{\mathbf{\hat{C}}_{ij}(t)\mathbf{\hat{B}}_{ij}(T)\mathbf{\hat{C}}_{ij}(t)}\Bigg\}
\end{equation}
\begin{equation}
Tr^{2}\bigg\{\sqrt{\beta_{ij}^{-2}\mathbf{\hat{P}}(\Delta_{i})\big|\psi_{j}(t)\big\rangle\big\langle \psi_{j}(t)\big|\mathbf{\hat{P}}(\Delta_{i})\mathbf{\hat{B}}_{ij}(T)\mathbf{\hat{P}}(\Delta_{i})\big|\psi_{j}(t)\big\rangle\big\langle \psi_{j}(t)\big|\mathbf{\hat{P}}(\Delta_{i})}\bigg\} = 
\end{equation}
\begin{equation}
\beta_{ij}^{-1}\big\langle \psi_{j}(t)\big|\mathbf{\hat{P}}(\Delta_{i})\mathbf{\hat{B}}_{ij}(T)\mathbf{\hat{P}}(\Delta_{i})\big|\psi_{j}(t)\big\rangle Tr^{2}\bigg\{\sqrt{\beta_{ij}^{-1}\mathbf{\hat{P}}(\Delta_{i})\big|\psi_{j}(t)\big\rangle\big\langle \psi_{j}(t)\big|\mathbf{\hat{P}}(\Delta_{i})}\bigg\} =   
\end{equation}
\begin{equation}
\beta_{ij}^{-1}\big\langle \psi_{j}(t)\big|\mathbf{\hat{P}}(\Delta_{i})\mathbf{\hat{B}}_{ij}(T)\mathbf{\hat{P}}(\Delta_{i})\big|\psi_{j}(t)\big\rangle = Tr\big\{\mathbf{\hat{C}}_{ij}(t)\mathbf{\hat{B}}_{ij}(T) \big\} 
\end{equation}

Recapitulating, 

\begin{equation}
\label{eqn:hanc}
\mathscr{D}_{\mathbf{\hat{A}}}\big(T,\Delta\big) \leq \bigg\langle\!\!\!\bigg\langle\sum_{j}\alpha_{j}\sum_{i}\beta_{ij}\Big(Tr^{2}\big\{ \mathbf{\hat{C}}_{ij}(t)+\mathbf{\hat{B}}_{ij}(T)\big\}-4Tr\big\{\mathbf{\hat{C}}_{ij}(t)\mathbf{\hat{B}}_{ij}(T) \big\}\Big)\bigg\rangle\!\!\!\bigg\rangle_{T} 
\end{equation}
The benefits of the latter estimate are the $\mathbf{\hat{A}}$-independence and its amenability to taking averages over the time domain $[0,T]$. It is clear from the above that $\mathscr{D}(T,\Delta)$ is small when the quantum map $\langle\!\langle \mathbf{\hat{U}}_{t}(\cdot) \mathbf{\hat{U}}^{\dagger}\rangle\!\rangle_{T}$ acts as a small perturbation to the identity map when acting on the $|\psi_{j}\rangle\langle \psi_{j}|$. In what follows, let $t^{*}$ be the $t$ which maximizes the term (\ref{eqn:hanc}). 
\begin{equation}
(\ref{eqn:hanc}) \leq 
\sup_{t\in[0,T]}\sum_{j}\alpha_{j}\sum_{i}\beta_{ij}\Big(Tr^{2}\big\{ \mathbf{\hat{C}}_{ij}(t)+\mathbf{\hat{B}}_{ij}(T)\big\}-4Tr\big\{\mathbf{\hat{C}}_{ij}(t)\mathbf{\hat{B}}_{ij}(T) \big\}\Big) =
\end{equation}
\begin{equation}
\sum_{j}\alpha_{j}\sum_{i}\beta_{ij}\Big(Tr^{2}\big\{ \mathbf{\hat{C}}_{ij}(t^{*})+\mathbf{\hat{B}}_{ij}(T)\big\}-4Tr\big\{\mathbf{\hat{C}}_{ij}(t^{*})\mathbf{\hat{B}}_{ij}(T) \big\}\Big) =
\end{equation}
\begin{equation}
\sum_{j}\alpha_{j}\sum_{i}\beta_{ij}\Big(1+2Tr\big\{ \mathbf{\hat{B}}_{ij}(T)\big\}+Tr^{2}\big\{ \mathbf{\hat{B}}_{ij}(T)\big\}-4Tr\big\{\mathbf{\hat{C}}_{ij}(t^{*})\mathbf{\hat{B}}_{ij}(T) \big\}\Big) 
\end{equation}
Moreover, 
\begin{equation}
Tr\big\{ \mathbf{\hat{B}}_{ij}(T)\big\} = \beta_{ij}^{-1}Tr\big\{\mathbf{\hat{P}}\big(\Delta_{i}\big)\Big\langle\!\!\Big\langle\big|\psi_{j}(t)\big\rangle\big\langle\psi_{j}(t)\big|\Big\rangle\!\!\Big\rangle_{T}\mathbf{\hat{P}}\big(\Delta_{i}\big)\big\}= 
\end{equation}
\begin{equation}
\beta_{ij}^{-1}\Big\langle\!\!\Big\langle Tr\big\{\mathbf{\hat{P}}\big(\Delta_{i}\big)\big|\psi_{j}(t)\big\rangle\big\langle\psi_{ij}(t)\big|\mathbf{\hat{P}}\big(\Delta_{i}\big)\big\}\Big\rangle\!\!\Big\rangle_{T} = \beta_{ij}^{-1}\Big\langle\!\!\Big\langle Tr\big\{\mathbf{\hat{P}}\big(\Delta_{i}\big)\big|\psi_{j}(0)\big\rangle\big\langle\psi_{ij}(0)\big|\mathbf{\hat{P}}\big(\Delta_{i}\big)\big\}\Big\rangle\!\!\Big\rangle_{T}=
\end{equation}
\begin{equation}
\beta_{ij}^{-1}Tr\big\{\mathbf{\hat{P}}\big(\Delta_{i}\big)\big|\psi_{j}(0)\big\rangle\big\langle\psi_{ij}(0)\big|\mathbf{\hat{P}}\big(\Delta_{i}\big)\big\} =
\end{equation}
\begin{equation}
\label{eqn:pen2}
\beta_{ij}^{-1}\beta_{ij} =1 
\end{equation}
Furthermore,
\begin{equation}
Tr\big\{\mathbf{\hat{C}}_{ij}(t^{*})\mathbf{\hat{B}}_{ij}(T) \big\} = \beta_{ij}^{-2}Tr\Big\{ \mathbf{\hat{P}}\big(\Delta_{i}\big)\big|\psi_{j}(t^{*})\big\rangle\big\langle\psi_{j}(t^{*})\big|\mathbf{\hat{P}}\big(\Delta_{i}\big)\Big\langle\!\!\!\Big\langle\big|\psi_{j}(t)\big\rangle\big\langle\psi_{j}(t)\big|\Big\rangle\!\!\!\Big\rangle_{T}\mathbf{\hat{P}}\big(\Delta_{i}\big) \Big\}  =  
\end{equation}
\begin{equation}
\beta_{ij}^{-2}\Big\langle\!\!\Big\langle\big|\big\langle\psi_{j}(t^{*})\big|\mathbf{\hat{P}}\big(\Delta_{i}\big)\big|\psi_{j}(t)\big\rangle\big|^{2}\Big\rangle\!\!\Big\rangle_{T} \geq \beta_{ij}^{-2}\big|\big\langle\psi_{j}(t^{*})\big|\mathbf{\hat{P}}\big(\Delta_{i}\big)\big|\psi_{j}(\gamma^{*})\big\rangle\big|^{2}=
\end{equation}
\begin{equation}
\label{eqn:pen1}
\beta_{ij}^{-2}\Big|\int_{\Delta_{i}}e^{-i(t^{*}-\gamma^{*})x}d\mu_{\psi_{j}}(x)\Big|^{2}
\end{equation}
where $\gamma^{*}$ minimizes $\big|\big\langle\psi_{j}(t^{*})\big|\mathbf{\hat{P}}\big(\Delta_{i}\big)\big|\psi_{j}(t)\big\rangle\big|^{2}$ over $t\in[0,T]$.
Now, using (\ref{eqn:pen1}) and (\ref{eqn:pen2}) we get the following estimate.
\begin{equation}
\mathscr{D}_{\mathbf{\hat{A}}}\big(T,\Delta\big) \leq  4\sum_{j}\alpha_{j}\sum_{i}\beta_{ij}\Bigg(1-\beta_{ij}^{-2}\Big|\int_{\Delta_{i}}e^{-i(t^{*}-\gamma^{*})x}\mu_{\psi_{j}}(x)\Big|^{2}\Bigg)\leq
\end{equation}

\begin{equation}
 4\max_{i,j}\Bigg(1-\beta_{ij}^{-2}\Big|\int_{\Delta_{i}}e^{-i(t^{*}-\gamma^{*})x}\mu_{\psi_{j}}(dx)\Big|^{2}\Bigg)\leq
\end{equation}
\begin{equation}
 4\max_{i,j}\sup_{\tau\in[-T,T]}\Bigg(1-\beta_{ij}^{-2}\Big|\int_{\Delta_{i}}e^{-i\tau x}\mu_{\psi_{j}}(dx)\Big|^{2}\Bigg)
\end{equation}
\end{proof}

\end{document}